\documentclass[aps,amsmath,amssymb,nofootinbib,twocolumn]{revtex4-1}
\usepackage{color}
\usepackage[colorlinks=true,linkcolor=blue,urlcolor=blue,citecolor=blue]{hyperref}
\usepackage{color}
\usepackage{times}

\usepackage{graphicx}

\newcommand{\hp}{\hphantom}

\usepackage{tikz}
\usetikzlibrary{arrows,shapes,positioning,decorations.pathmorphing}
\usetikzlibrary{decorations.markings}
\usetikzlibrary{snakes}
\tikzset{snake it/.style={decorate, decoration=snake,segment length=3mm}}
\tikzstyle arrowstyle=[scale=1]
\tikzstyle directed=[postaction={decorate,decoration={markings,
    mark=at position .65 with {\arrow[arrowstyle]{stealth}}}}]
\tikzstyle endreversedirected=[postaction={decorate,decoration={markings,
    mark=at position 1.0 with {\arrow[arrowstyle]{stealth}}}}]
\tikzstyle enddirected=[postaction={decorate,decoration={markings,
    mark=at position 1.0 with {\arrow[arrowstyle]{stealth}}}}]
\tikzstyle reverse directed=[postaction={decorate,decoration={markings,
    mark=at position .65 with {\arrowreversed[arrowstyle]{stealth};}}}]

\tikzset{cross/.style={cross out, draw=black, minimum size=2*(#1-\pgflinewidth), inner sep=0pt, outer sep=0pt},
cross/.default={1pt}}

\newcommand{\Ione}{{\parbox{1.15cm}{{\begin{tikzpicture}
\coordinate (x1) at  (0.35,0) ; 
\coordinate (x2) at  (1.35,0) ; 
\fill (x1) circle (2pt);
\fill (x2) circle (2pt);
\draw   (x2) arc(60:120:1);
\draw   (x2) arc(-60:-120:1);
\end{tikzpicture}}}}}

\newcommand{\Ibanana}{{\parbox{2.15cm}{{\begin{tikzpicture}
\coordinate (x1) at  (0.35,0) ; 
\coordinate (x2) at  (1.35,0) ;
\coordinate (x3) at  (2.35,0) ; 
\fill (x1) circle (2pt);
\fill (x2) circle (2pt);
\fill (x3) circle (2pt);
\draw   (x2) arc(60:120:1);
\draw   (x2) arc(-60:-120:1);
\draw   (x3) arc(60:120:1);
\draw   (x3) arc(-60:-120:1);
\end{tikzpicture}}}}}

\newcommand{\ILH}{{\parbox{1.15cm}{{\begin{tikzpicture}
\coordinate (x1) at  (0.35,0) ; 
\coordinate (x2) at  (1.35,0) ; 
\coordinate (x3) at  (0.85,0.85) ; 
\fill (x1) circle (2pt);
\fill (x2) circle (2pt);
\fill (x3) circle (2pt);
\draw (x1) -- (x3) ;
\draw   (x2) arc(60:120:1);
\draw   (x2) arc(-60:-120:1);
\draw   (x3) -- (x2);
\end{tikzpicture}}}}}

\newcommand{\Isunset}{{\parbox{1.15cm}{{\begin{tikzpicture}
\coordinate (x1) at  (0.3,0) ; 
\coordinate (x2) at  (1.3,0) ; 
\fill (x1) circle (2pt);
\fill (x2) circle (2pt);
\draw  (x2) arc(30:150:0.58);
\draw  (x2) arc(-30:-150:0.58);
\draw(x1)--(x2);
\end{tikzpicture}}}}}

\newcommand{\Ii}{{\parbox{1.02cm}{{\begin{tikzpicture}
\coordinate (x1) at  (-0.433013,-0.25 ) ; 
\coordinate (x2) at  (0.433013,-0.25 ) ; 
\coordinate (x3) at  (0,0.5) ; 
\coordinate (x4) at  (0,0) ; 
\fill (x1) circle (2pt);
\fill (x2) circle (2pt);
\fill (x3) circle (2pt);
\fill (x4) circle (2pt);
\draw  (x3) arc(90:-270:0.5);
\draw (x1)--(x4);
\draw (x2)--(x4);
\draw (x3)--(x4);
\end{tikzpicture}}}}}

\newcommand{\ITP}{{\parbox{0.42cm}{{\begin{tikzpicture}
\coordinate (x1) at  (0,0) ; 
\fill (x1) circle (2pt);
\draw  (x1) arc(-90:270:0.2);
\end{tikzpicture}}}}}

\newcommand{\ItwoloopTP}{{\parbox{1.5cm}{{\begin{tikzpicture}
\coordinate (x1) at  (0.35,0) ; 
\coordinate (x2) at  (1.35,0) ; 
\fill (x1) circle (2pt);
\fill (x2) circle (2pt);
\draw   (x2) arc(60:120:1);
\draw   (x2) arc(-60:-120:1);
\draw  (x1) arc(0:360:0.2);
\end{tikzpicture}}}}}

\newcommand{\Ir}{{\parbox{1.15cm}{{\begin{tikzpicture}
\coordinate (x1) at  (0.35,0) ; 
\coordinate (x2) at  (1.35,0) ; 
\coordinate (x3) at  (0.85,0.85) ; 
\fill (x1) circle (2pt);
\fill (x2) circle (2pt);
\fill (x3) circle (2pt);
\draw   (x2) arc(60:120:1);
\draw   (x2) arc(-60:-120:1);
\draw   (x1) arc(-60:0:1);
\draw   (x1) arc(180:120:1);
\draw   (x3) -- (x2);
\end{tikzpicture}}}}}

\newcommand{\Iss}{{\parbox{1.15cm}{{\begin{tikzpicture}
\coordinate (x1) at  (0.35,0) ; 
\coordinate (x2) at  (1.35,0) ; 
\fill (x1) circle (2pt);
\fill (x2) circle (2pt);
\draw   (x2) arc(60:120:1);
\draw   (x2) arc(-60:-120:1);
\draw   (x1) -- (x2);
\end{tikzpicture}}}}}

\newcommand{\IrB}{{\parbox{1.15cm}{{\begin{tikzpicture}
\coordinate (x1) at  (0.35,0) ; 
\coordinate (x2) at  (1.35,0) ; 
\coordinate (x3) at  (0.85,0.85) ; 
\coordinate (x4) at  (0.45,0.425);
\fill (x1) circle (2pt);
\fill (x2) circle (2pt);
\fill (x3) circle (2pt);
\draw   (x2) arc(60:120:1);
\draw   (x2) arc(-60:-120:1);
\draw   (x1) arc(-60:0:1);
\draw   (x1) arc(180:120:1);
\draw   (x3) -- (x2);
\draw (x4) circle (2pt);
\end{tikzpicture}}}}}

\newcommand{\IrA}{{\parbox{1.15cm}{{\begin{tikzpicture}
\coordinate (x1) at  (0.35,0) ; 
\coordinate (x2) at  (1.35,0) ; 
\coordinate (x3) at  (0.85,0.85) ; 
\coordinate (x4) at  (1.1,0.425);
\fill (x1) circle (2pt);
\fill (x2) circle (2pt);
\fill (x3) circle (2pt);
\draw   (x2) arc(60:120:1);
\draw   (x2) arc(-60:-120:1);
\draw   (x1) arc(-60:0:1);
\draw   (x1) arc(180:120:1);
\draw   (x3) -- (x2);
\draw (x4) circle (2pt);
\end{tikzpicture}}}}}

\newcommand{\Is}{{\parbox{1.15cm}{{\begin{tikzpicture}
\coordinate (x1) at  (0.35,0) ; 
\coordinate (x2) at  (1.35,0) ; 
\coordinate (x3) at  (0.85,0.85) ; 
\fill (x1) circle (2pt);
\fill (x2) circle (2pt);
\fill (x3) circle (2pt);
\draw (x1) -- (x3) ;
\draw   (x2) arc(60:120:1);
\draw   (x2) arc(-60:-120:1);
\draw   (x3) -- (x2);
\draw  (x3) arc(-90:270:0.2);
\end{tikzpicture}}}}}

\newcommand{\It}{{\parbox{1.15cm}{{\begin{tikzpicture}
\coordinate (x1) at  (0.35,0) ; 
\coordinate (x2) at  (1.35,0) ; 
\coordinate (x3) at  (0.85,0.85) ; 
\fill (x1) circle (2pt);
\fill (x2) circle (2pt);
\fill (x3) circle (2pt);
\draw   (x2) arc(60:120:1);
\draw   (x2) arc(-60:-120:1);
\draw   (x1) arc(-60:0:1);
\draw   (x1) arc(180:120:1);
\draw  (x3) arc(-90:270:0.2);
\end{tikzpicture}}}}}

\newcommand{\Iu}{{\parbox{1.15cm}{{\begin{tikzpicture}
\coordinate (x1) at  (0.35,0) ; 
\coordinate (x2) at  (1.35,0) ; 
\coordinate (x3) at  (0.85,0.85) ; 
\fill (x1) circle (2pt);
\fill (x2) circle (2pt);
\fill (x3) circle (2pt);
\draw   (x2) arc(60:120:1);
\draw   (x2) arc(-60:-120:1);
\draw (x1) -- (x2) -- (x3) -- (x1) ;
\end{tikzpicture}}}}}

\newcommand{\Iv}{{\parbox{1.8cm}{{\begin{tikzpicture}
\coordinate (x1) at  (0.35,0) ; 
\coordinate (x2) at  (1.35,0) ; 
\coordinate (x3) at  (0.85,0.85) ; 
\fill (x1) circle (2pt);
\fill (x2) circle (2pt);
\fill (x3) circle (2pt);
\draw   (x3) -- (x2) -- (x1) -- (x3);
\draw  (x1) arc(0:360:0.2);
\draw  (x2) arc(-180:180:0.2);
\end{tikzpicture}}}}}

\newcommand{\Iw}{{\parbox{1.15cm}{{\begin{tikzpicture}
\coordinate (x1) at  (0.35,0) ; 
\coordinate (x2) at  (1.35,0) ; 
\coordinate (x3) at  (0.85,0.85) ; 
\fill (x1) circle (2pt);
\fill (x2) circle (2pt);
\fill (x3) circle (2pt);
\draw   (x2) arc(60:120:1);
\draw   (x2) arc(-60:-120:1);
\draw   (x1) arc(-60:0:1);
\draw   (x1) arc(180:120:1);
\draw  (x1) -- (x2) ;
\end{tikzpicture}}}}}

\newcommand{\Ix}{{\parbox{1.15cm}{{\begin{tikzpicture}
\coordinate (x1) at  (0.35,0) ; 
\coordinate (x2) at  (1.35,0) ; 
\coordinate (x3) at  (0.85,0.85) ; 
\fill (x1) circle (2pt);
\fill (x2) circle (2pt);
\fill (x3) circle (2pt);
\draw   (x1) arc(-60:0:1);
\draw   (x1) arc(180:120:1);
\draw   (x3) -- (x2) -- (x1);
\draw  (x3) arc(-90:270:0.2);
\end{tikzpicture}}}}}

\newcommand{\Iy}{{\parbox{1.15cm}{{\begin{tikzpicture}
\coordinate (x1) at  (0.35,0) ; 
\coordinate (x2) at  (1.35,0) ; 
\coordinate (x3) at  (0.85,0.85) ; 
\fill (x1) circle (2pt);
\fill (x2) circle (2pt);
\fill (x3) circle (2pt);
\draw   (x1) arc(-60:0:1);
\draw   (x1) arc(180:120:1);
\draw (x3) arc(-180:-120:1);
\draw   (x3) arc(60:0:1);
\draw  (x3) arc(-90:270:0.2);
\end{tikzpicture}}}}}

\newcommand{\Iz}{{\parbox{1.15cm}{{\begin{tikzpicture}
\coordinate (x1) at  (0.35,0) ; 
\coordinate (x2) at  (1.35,0) ; 
\coordinate (x3) at  (0.85,0.85) ; 
\fill (x1) circle (2pt);
\fill (x2) circle (2pt);
\fill (x3) circle (2pt);
\draw   (x2) -- (x1);
\draw   (x1) arc(-60:0:1);
\draw (x3) arc(-180:-120:1);
\draw  (x3) arc(0:360:0.2);
\draw  (x3) arc(-180:180:0.2);
\end{tikzpicture}}}}}

\newcommand{\IfourA}{{\parbox{1.35cm}{{\begin{tikzpicture}
\coordinate (x1) at  (0.35,0) ; 
\coordinate (x2) at  (1.35,0) ; 
\coordinate (x3) at  (0.35,1) ; 
\coordinate (x4) at  (1.35,1) ; 
\fill (x1) circle (2pt);
\fill (x2) circle (2pt);
\fill (x3) circle (2pt);
\fill (x4) circle (2pt);
\draw   (x3) arc(150:210:1);
\draw  (x3) arc(30:-30:1);
\draw  (x4) arc(150:210:1);
\draw  (x4) arc(30:-30:1);
\draw  (x1) -- (x2) -- (x3) -- (x4);
\end{tikzpicture}}}}}

\newcommand{\IfourB}{{\parbox{1.35cm}{{\begin{tikzpicture}
\coordinate (x1) at  (0.35,0) ; 
\coordinate (x2) at  (1.35,0) ; 
\coordinate (x3) at  (0.35,1) ; 
\coordinate (x4) at  (1.35,1) ; 
\fill (x1) circle (2pt);
\fill (x2) circle (2pt);
\fill (x3) circle (2pt);
\fill (x4) circle (2pt);
\draw   (x4) arc(60:120:1);
\draw  (x4) arc(-60:-120:1);
\draw (x1) -- (x2) ;
\draw   (x3) arc(150:210:1);
\draw  (x3) arc(30:-30:1);
\draw (x4) arc(150:210:1);
\draw   (x4) arc(30:-30:1);
\end{tikzpicture}}}}}

\newcommand{\IfourC}{{\parbox{1.35cm}{{\begin{tikzpicture}
\coordinate (x1) at  (0.85,0.5) ; 
\coordinate (x2) at  (1.35,0) ; 
\coordinate (x3) at  (0.35,1) ; 
\coordinate (x4) at  (1.35,1) ; 
\fill (x1) circle (2pt);
\fill (x2) circle (2pt);
\fill (x3) circle (2pt);
\fill (x4) circle (2pt);
\draw  (x4) arc(60:120:1);
\draw   (x4) arc(-60:-120:1);
\draw  (x2) arc(-90:-180:1);
\draw  (x3) --(x1) -- (x2) -- (x4) -- (x1) ; 
\end{tikzpicture}}}}}

\newcommand{\Ih}{{\parbox{1.28cm}{{\begin{tikzpicture}
\node (x1) at  (0.35,0) {} ; 
\node (x2) at  (1.35,0) {}; 
\node (x3) at  (0.35,1) {}; 
\node (x4) at  (1.35,1) {}; 
\fill (x1) circle (2pt);
\fill (x2) circle (2pt);
\fill (x3) circle (2pt);
\fill (x4) circle (2pt);
\draw  (x4) arc(60:120:1);
\draw   (x4) arc(-60:-120:1);
\draw   (x2) arc(60:120:1);
\draw   (x2) arc(-60:-120:1);
\draw  (x3) arc(150:210:1);
\draw  (x3) arc(30:-30:1);
\end{tikzpicture}}}}}

\newcommand{\Ij}{{\parbox{1.15cm}{{\begin{tikzpicture}
\coordinate (x1) at  (0.35,0) ; 
\coordinate (x2) at  (1.35,0) ; 
\coordinate (x3) at  (0.35,1) ; 
\coordinate (x4) at  (1.35,1) ; 
\fill (x1) circle (2pt);
\fill (x2) circle (2pt);
\fill (x3) circle (2pt);
\fill (x4) circle (2pt);
\draw  (x2) arc(-115:-155:2);
\draw  (x2) arc(25:65:2);
\draw  (x1) -- (x2) -- (x4) -- (x3) --(x1); 
\end{tikzpicture}}}}}

\newcommand{\Ik}{{\parbox{1.28cm}{{\begin{tikzpicture}
\coordinate (x1) at  (0.35,0); 
\coordinate (x2) at  (1.35,0) ; 
\coordinate (x3) at  (0.35,1); 
\coordinate (x4) at  (1.35,1) ; 
\fill (x1) circle (2pt);
\fill (x2) circle (2pt);
\fill (x3) circle (2pt);
\fill (x4) circle (2pt);
\draw  (x3) arc(150:210:1);
\draw  (x3) arc(30:-30:1);
\draw (x1)--(x3)--(x4)--(x2)--(x1);
\end{tikzpicture}}}}}

\newcommand{\Il}{{\parbox{1.28cm}{{\begin{tikzpicture}
\coordinate (x1) at  (0.35,0); 
\coordinate (x2) at  (1.35,0) ; 
\coordinate (x3) at  (0.35,1); 
\coordinate (x4) at  (1.35,1) ; 
\fill (x1) circle (2pt);
\fill (x2) circle (2pt);
\fill (x3) circle (2pt);
\fill (x4) circle (2pt);
\draw  (x3) arc(150:210:1);
\draw  (x3) arc(30:-30:1);
\draw (x3)--(x4)--(x2)--(x1);
\draw (x4)--(x1);
\end{tikzpicture}}}}}

\newcommand{\IIm}{{\parbox{1.35cm}{{\begin{tikzpicture}
\coordinate (x1) at  (0.35,0) ; 
\coordinate (x2) at  (1.35,0) ; 
\coordinate (x3) at  (0.35,1) ; 
\coordinate (x4) at  (1.35,1) ; 
\fill (x1) circle (2pt);
\fill (x2) circle (2pt);
\fill (x3) circle (2pt);
\fill (x4) circle (2pt);
\draw (x1) -- (x2) ;
\draw (x3) -- (x4) ;
\draw   (x3) arc(150:210:1);
\draw  (x3) arc(30:-30:1);
\draw (x4) arc(150:210:1);
\draw   (x4) arc(30:-30:1);
\end{tikzpicture}}}}}

\newcommand{\In}{{\parbox{1.25cm}{{\begin{tikzpicture}
\coordinate (x1) at  (0.35,0) ; 
\coordinate (x2) at  (1.35,0) ; 
\coordinate (x3) at  (0.35,1) ; 
\coordinate (x4) at  (1.35,1) ; 
\fill (x1) circle (2pt);
\fill (x2) circle (2pt);
\fill (x3) circle (2pt);
\fill (x4) circle (2pt);
\draw  (x2) arc(-115:-155:2);
\draw  (x2) arc(25:65:2);
\draw   (x2) arc(60:120:1);
\draw   (x2) arc(-60:-120:1);
\draw (x4) arc(150:210:1);
\draw   (x4) arc(30:-30:1);
\end{tikzpicture}}}}}

\newcommand{\Io}{{\parbox{1.25cm}{{\begin{tikzpicture}
\coordinate (x1) at  (0.35,0) ; 
\coordinate (x2) at  (1.35,0) ; 
\coordinate (x3) at  (0.35,1) ; 
\coordinate (x4) at  (1.35,1) ; 
\fill (x1) circle (2pt);
\fill (x2) circle (2pt);
\fill (x3) circle (2pt);
\fill (x4) circle (2pt);
\draw  (x1) -- (x3) -- (x4) ; 
\draw   (x2) arc(60:120:1);
\draw   (x2) arc(-60:-120:1);
\draw (x4) arc(150:210:1);
\draw   (x4) arc(30:-30:1);
\end{tikzpicture}}}}}

\newcommand{\Ip}{{\parbox{1.25cm}{{\begin{tikzpicture}
\coordinate (x1) at  (0.35,0) ; 
\coordinate (x2) at  (1.35,0) ; 
\coordinate (x3) at  (0.35,1) ; 
\coordinate (x4) at  (1.35,1) ; 
\fill (x1) circle (2pt);
\fill (x2) circle (2pt);
\fill (x3) circle (2pt);
\fill (x4) circle (2pt);
\draw  (x2) arc(-115:-155:2);
\draw  (x2) arc(25:65:2);
\draw   (x2) -- (x1) -- (x3) ; 
\draw (x4) arc(150:210:1);
\draw   (x4) arc(30:-30:1);
\end{tikzpicture}}}}}

\newcommand{\Iq}{{\parbox{1.35cm}{{\begin{tikzpicture}
\coordinate (x1) at  (0.35,0) ; 
\coordinate (x2) at  (1.35,0) ; 
\coordinate (x3) at  (0.35,1) ; 
\coordinate (x4) at  (1.35,1) ; 
\fill (x1) circle (2pt);
\fill (x2) circle (2pt);
\fill (x3) circle (2pt);
\fill (x4) circle (2pt);
\draw   (x2) -- (x1) -- (x4) ; 
\draw (x4) arc(150:210:1);
\draw   (x4) arc(30:-30:1);
\draw (x3) arc(150:210:1);
\draw   (x3) arc(30:-30:1);
\end{tikzpicture}}}}}

\definecolor{labelkey}{cmyk}{.4,.2,0,0}

\usepackage{graphicx}

\newcommand{\fig}[2]{\includegraphics[width=#1]{./#2}}

\newcommand{\Fig}[1]{\includegraphics[width=\columnwidth]{./#1}}
\newlength{\bilderlength}

\newcommand{\half}{\frac{1}{2}}

\begin{document}
\newcommand{\Eq}[1]{Eq.~(\ref{#1})}
\newcommand{\Eqs}[1]{Eqs.~(\ref{#1})}
\newcommand{\eq}[1]{(\ref{#1})}
\newcommand{\ds}[1]{\displaystyle }
\newcommand{\bra}[1]{\left<#1\right|}
\newcommand{\ket}[1]{\left|#1\right>}
\newcommand{\braket}[2]{\left.\left<#1\right|#2\right>}
\newcommand{\blue}{\color{blue}}
\newcommand{\bea}{\begin{eqnarray}}
\newcommand{\eea}{\end{eqnarray}}
\newcommand{\be}{\begin{equation}}
\newcommand{\ee}{\end{equation}}
\newcommand{\red}{\color{red}}
\newcommand{\black}{\color{black}}

\newcommand{\ca}[1]{{\cal #1}}
\newcommand{\sgn}{{\mathrm{sgn}}}
\newcommand{\rme}{{\mathrm{e}}}
\newcommand{\rmd}{{\mathrm{d}}} 
\newcommand{\nn}{\nonumber}
\newcommand{\E}{\epsilon}

\newcommand{\PLDKW}{\cite{LeDoussalWieseChauve2002}}

%
\title{\sffamily\bfseries\large Roughness and critical force for depinning at 3-loop order}
\author{\sffamily\bfseries\normalsize  Mikhail N. Semeikin and Kay
J\"org Wiese}
\affiliation{CNRS-Laboratoire de Physique de l'Ecole Normale Sup\'erieure, PSL, ENS, Sorbonne Universit\'e, Universit\'e Paris Cit\'e, 24 rue Lhomond, 75005 Paris, France}

\begin{abstract}
A $d$-dimensional elastic manifold  at depinning is described by a renormalized field theory, based on the Functional Renormalization Group (FRG). 
Here we analyze this theory to 3-loop order, equivalent to third order in $\epsilon=4-d$, where $d$ is the internal dimension. 
The critical exponent reads $\zeta = \frac \epsilon3 + 0.04777 \epsilon^2 -0.068354 \epsilon^3 + \ca O(\epsilon^4)$. Using that $\zeta(d=0)=2^-$, we estimate $\zeta(d=1)=1.266(20)$, $\zeta(d=2)=0.752(1)$ and $\zeta(d=3)=0.357(1)$. For Gaussian disorder, the   pinning force per site is estimated as $f_{\rm c}= \ca B m^{2}\rho_m + f_{\rm c}^0$, where $m^2$ is the strength of the confining potential, $\ca B$ a universal amplitude, $\rho_m$ the correlation length of the disorder, and $f_{\rm c}^0$ a non-universal lattice  dependent term.  For charge-density waves, we  find a  mapping to the standard $\phi^4$-theory with $O(n)$ 
symmetry in the limit of $n\to -2$. This gives $f_{\rm c} =  \tilde {\ca A}(d) m^2 \ln (m) + f_{\rm c}^0 $, with $\tilde {\ca A}(d) = -\partial_n \big[\nu(d,n)^{-1}+\eta(d,n)\big]_{n=-2}$, reminiscent of log-CFTs.
\end{abstract}

\maketitle

\section{Introduction}
Many disordered elastic systems undergo a depinning transition. Examples are  magnetic domain walls \cite{AlessandroBeatriceBertottiMontorsi1990,DurinZapperi2000,HuthHaibachAdrian2002,CerrutiDurinZapperi2009,JeudyMouginBustingorrySavero-TorresGorchonKoltonLemaitreJamet2016,DurinBohnCorreaSommerDoussalWiese2016,JeudyDiaz-PardoSavero-TorresBustingorryKolton2018,terBurgBohnDurinSommerWiese2021}, earthquakes \cite{GutenbergRichter1944,GutenbergRichter1956,BenZionRice1993,CarlsonLangerShaw1994,FisherDahmenRamanathanBenZion1997,DSFisher1998,Kagan2002,JaglaKolton2009}, contact lines 
\cite{BrochardGennes1991,AmaralBarabasiStanley1994,RolleyGuthmannGombrowiczRepain1998,PrevostRolleyGuthmann2002,MoulinetRossoKrauthRolley2004,RolleyGuthmann2007,RossoKrauth2002,BachasLeDoussalWiese2006}, vortex lattices \cite{ScheidlVinokur1998,BasslerPaczuski1998,LeDoussalRistivojevicWiese2013,EmigNattermann2006,GiamarchiLeDoussal1995}, charge-density waves \cite{WieseFedorenko2019,ScheidlVinokur1998,DSFisher1998,EmigNattermann1997,ChenBalentsFisherMarchetti1996}, and many more, see the recent review \cite{Wiese2021}. 

They all evolve via an overdamped Langevin equation for the position $u(x,t)$ of site $x$ at time $t$, 
\be\label{EOM}
\eta \partial_t u(x,t) = \nabla^2 u(x,t) + m^2 \left[ w- u(x,t) \right] + F\big( x, u(x,t) \big).
\ee
The second term on the r.h.s.\ stems from a confining potential of strength $m^2$, centered at $w$. Increasing $w$ adiabatically slowly drives the system. 
The last term $F(x,u)$ is a short-range correlated random force,  possibly the $u$-derivative of a random potential. It is assumed to be Gaussian with variance
(connected part)
\be\label{Delta-def}
\overline{F(x,u) F(x',u')}^{\rm c} = \delta^d(x-x') \Delta_0(u-u').
\ee 
The overbar denotes a disorder average. 

The field theory of depinning is by now well established (see the review  \cite{Wiese2021}).
It relies on a functional renormalization group for the disorder correlator $\Delta(u)$, starting from the microscopic disorder $\Delta_0(u)$. This idea, already present in the seminal works of Wilson \cite{WilsonKogut1974} and Wegner\&Houghton \cite{WegnerHoughton1973} was recognized as crucial by D.~Fisher in collaboration with   Narayan and Balents, \cite{Fisher1985b,DSFisher1985,DSFisher1986,MiddletonFisher1991,NarayanDSFisher1992b,NarayanDSFisher1992a,NarayanDSFisher1993a,BalentsDSFisher1993}, as well as Leschhorn, Nattermann, Stepanow and Tang \cite{NattermannStepanowTangLeschhorn1992,LeschhornNattermannStepanowTang1997}.
Later, Chauve, Le Doussal and Wiese \cite{ChauveLeDoussalWiese2000a,LeDoussalWieseChauve2002,LeDoussalWieseChauve2003}
showed that both in equilibrium and at depinning a consistent field theory exists up to 2-loop order. This field theory allows us to deal with the many non-trivial observables arising for pinned manifolds, and especially to treat quantitatively avalanches \cite{LeDoussalWiese2008c,LeDoussalMiddletonWiese2008,RossoLeDoussalWiese2009a,LeDoussalWiese2011b,LeDoussalWiese2010b,DobrinevskiLeDoussalWiese2011b,LeDoussalWiese2011a,LeDoussalPetkovicWiese2012,DobrinevskiLeDoussalWiese2013,LeDoussalWiese2012a,DobrinevskiLeDoussalWiese2014a}, including their distributions of size, velocity and shape in good agreement with simulations \cite{LeDoussalMiddletonWiese2008,RossoLeDoussalWiese2009a,ZhuWiese2017} and experiments \cite{DurinBohnCorreaSommerDoussalWiese2016}. 

Systems at depinning are characterized by a jerky motion for its center of mass $u_w$, 
\be\label{COM}
u_w := \frac1 {L^d} \int \rmd^d x\, u(x,t).
\ee
Here $L$ is the size of the system, and the integral is evaluated once all motion has stopped. 
The index $w$ refers to the position of the confining potential, which is adiabatically slowly moved forward. 
The central ingredient of the field theory is the renormalized force correlator, defined by the connected average
\be\label{FRG-Delta-def}
\Delta(w-w') :=m^4 L^d \overline{ (w-u_w)   (w'-u_{w'})}^{\rm c}.
\ee
On one hand, it can be calculated in a loop expansion, equivalent to an expansion in $\epsilon=4-d$, where $d$ is the internal dimension of the manifold. On the other hand, the prescription \eq{FRG-Delta-def} can be tested in simulations 
\cite{MiddletonLeDoussalWiese2006,RossoLeDoussalWiese2006a} and experiments \cite{LeDoussalWieseMoulinetRolley2009,WieseBercyMelkonyanBizebard2019,terBurgBohnDurinSommerWiese2021,terBurgRissoneRicoPastoRitortWiese2023}. 
For equilibrium, the loop expansion was  extended   to 3-loop order in Refs.~\cite{WieseHusemannLeDoussal2018,HusemannWiese2017}. 
Here we report 3-loop results for the $\beta$-function and the critical force at depinning. Our first central result is 
the roughness exponent $\zeta$
\be
 \zeta = \frac \epsilon3 + 0.0477709715 \epsilon^2 -0.0683544 (2) \epsilon^3 + \ca O(\epsilon^4).
\ee 
\begin{figure}[b]
\includegraphics[width=\columnwidth]{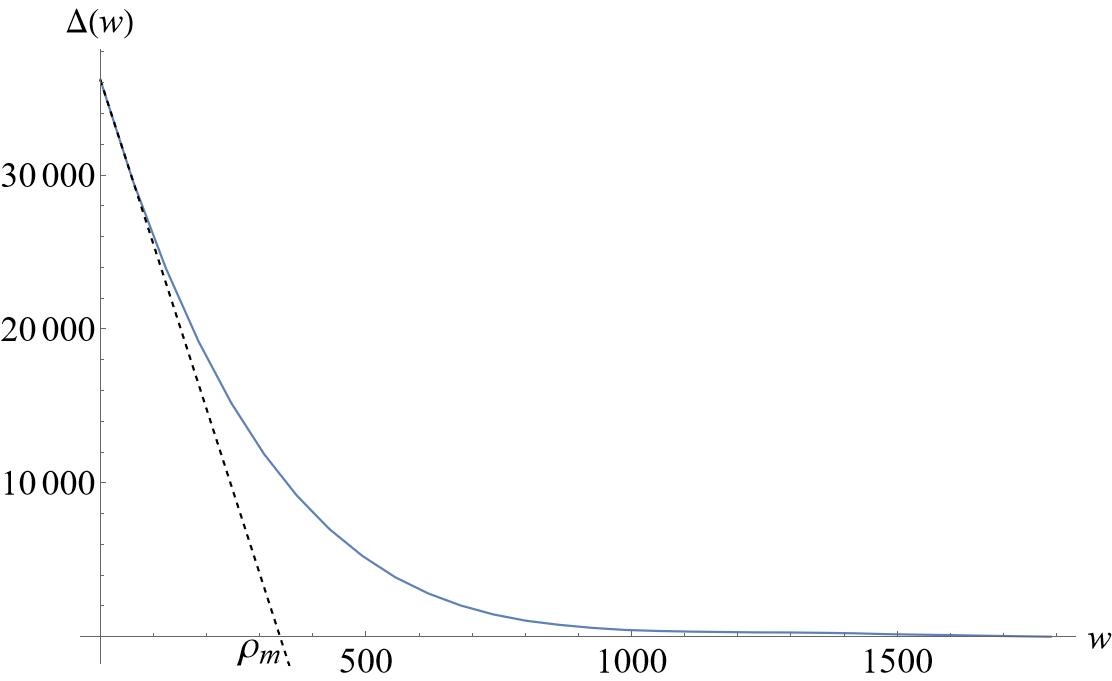}
\caption{$\Delta(w)$ for $mL = 6$ (blue) and $L=1024$. The tangent at $w=0$ defines the correlation length $\rho_m$.}
\label{fig:shape-example}
\end{figure} 
It can  numerically be measured by considering the finite-size scaling of the 2-point function in the limit of $m\to 0$,
\be
\int_{x,y} \overline{ [u(x)-u(y)]^2 }\sim L^{2d+2\zeta},
\ee
or for a finite $m$, 
\be
\overline{ [u(x)-u(y)]^2 }\sim m^{-2 \zeta} \mbox{ for }~ m|x-y| \gg 1.
\ee
The second relevant observable is the critical force per site (force density), 
\be
f_{\rm c} := m^2 \overline{(w -u_w)} \equiv -\overline {F(x,u(x,t))}.
\ee 
The last equality is verified by integrating the equation of motion \eq{EOM}, assuming periodic boundary conditions. 
We   show below that  
\bea\label{fc-1a}
 {f_{\rm c }}  &=& m^2 \overline {(w-u_w)} = f_{\rm c}^0 -  \ca B m^2 \rho_m ,  \\
\ca B &=&  1-0.30998 \epsilon +0.570136 \epsilon^2+\ca O (\epsilon ^3 ).\qquad 
\label{fc-1b}
\eea
Here $f_{\rm c}^0$ has a lattice-dependent (but $m$-independent) value, $\ca B$ is a universal amplitude, $m^2$ is given by the experiment,  and $\rho_m$ is a length scale in the driving direction set by the disorder, 
\be
\rho_m := \frac{\Delta(0)}{|\Delta'(0^+)|} \simeq \tilde \rho   m^{-\zeta}.
\ee 
An example for $\Delta(w)$ is given on Fig.~\ref{fig:shape-mL=4-versus-6}.
For small $m$, $f_{\rm c}$   converges to $m^{2-\zeta}$ times a system-specific amplitude $\tilde \rho$, set by the   microscopic disorder. 
While $\rho_m$ needs to be measured, the confining potential strength $m^2$ is imposed in the simulation or experiment. 
Should it be unknown, or insufficiently well known,   in the  experiment\footnote{This happens e.g.\ in DNA peeling or unzipping experiments, where  fluctuations of the beed diameter used in the optical trap induce fluctuations in $m^2$.}, it can be extracted from the linear part of the force-extension curve.

The amplitude $\ca B$ in \Eq{fc-1b} is universal, independent of   microscopic details. This is a quite astonishing result, as it is rarely possible to have a universal amplitude rather than a universal exponent. As we will see below, the reason this happens here is that the diagrammatic result does not resum into a power-law in $m$, but instead it depends logarithmically on $m$, 
and varying $\ca B \ln m$ gives back $\ca B$. The final result contains a power law, since it is still multiplied by $\rho_m$. We explain this in detail in section \ref{s:The critical force}.

The situation is even more extraordinary for charge-density waves (CDWs), for which $\zeta=0$. As we discuss in section  \ref{Critical force for CDWs}, CDWs can be mapped onto the $O(n)$ model in the limit of $n\to -2$. 
For this case, \Eqs{fc-1a}-\eq{fc-1b} reduce to 
\bea
\label{fc-2a}
\frac{f_{\rm c}}{m^2} &=&  \mbox{const}-\tilde {\ca A}(d)  \ln (m) , \\
\tilde {\ca A}(d) &=& -\partial_n \big[\nu(d,n)^{-1}+\eta(d,n)\big]_{n=-2}.
\label{fc-2b}
\eea
Here, $\nu(d,n)$ and $\eta(d,n)$ are the critical exponents $\nu$ and $\eta$ of the $O(n)$-model in dimension $d=4-\epsilon$.
\Eqs{fc-2a} and \eq{fc-2b} are  reminiscent  of log-CFTs \cite{Cardy2013,Cardy1999}:  when two operators collide as a function of an external control parameter, here $n$, the RG flow   becomes non-diagonalizable, and replaced by    a rank-2 Jordan-block form, leading to a universal amplitude in front of a logarithm, very much as in \Eqs{fc-2a}-\eq{fc-2b}.

The remainder of this article is organized as follows:
In section \ref{Renormalization group analysis} we start with the renormalization-group analysis for depinning. 
After a brief reminder of how to perform a functional RG, and the problems involved, we 
derive and analyze in section \ref{The beta-function and its fixed point} the RG $\beta$-function, the critical exponent $\zeta$, and the shape of the renormalized disorder correlator. 
The critical force is treated in section \ref{s:The critical force}. We then specialize to charge-density waves in section \ref{Critical force for CDWs}, which allows us to use high-order RG calculations for the $O(n)$-model. Section \ref{s:simulations} confirms our analytical calculations with numerical simulations. Conclusions are offered in section \ref{Conclusions}. 
Technical details and results for specific cases are given in various appendices. A table of contents can be found on page \pageref{tableofcontents}.

\section{Renormalization group analysis}
\label{Renormalization group analysis}

\subsection{Field theory of the depinning transition, response function}
Here we briefly review the basics of perturbation theory and renormalization for depinning as written in \Eq{EOM}.
For a   detailed introduction we refer the reader to section 3  of Ref.~\cite{Wiese2021}.
The idea is to discretize the equation of motion \eq{EOM}  as 
\bea\label{14}
&&\!\!\! u(x,t+\delta t) = u(x,t)  \nn\\
&&+ \frac{\delta t}{\eta} \Big[ \nabla^2 u(x,t) + m^2 \left[ w- u(x,t) \right] + F\big( x, u(x,t) \big)\Big].
\qquad 
\eea
For $x$ and $t$ fixed, this is achieved by writing, for the expectation $\langle \ca O\rangle$  of any observable  $\ca O$  depending on $u(x,t+\delta t)$ \cite{MSR,Janssen1976,DeDominicis1976,Janssen1985,Tauber2012,Vasilev2004}
\begin{widetext}
\bea\label{15}
\langle \ca O(u(x,t+\delta t))\rangle &=&\frac{\eta}{\delta t} \int\limits_{-i \infty}^{i \infty} \frac{\rmd \tilde u(x,t)}{2\pi i}\int\limits_{-\infty}^\infty \rmd u(x,t+\delta t)\, \ca O(u(x,t+\delta t)) \nn\\
&& \qquad \times \rme^{\tilde u(x,t)\left[\frac{\eta}{\delta t}\big( u(x,t+\delta t)- u(x,t) \big)+ \nabla^2 u(x,t) + m^2 \left[ w- u(x,t) \right] + F\big( x, u(x,t) \big)\right] }
\eea
\end{widetext}
The integral over $\tilde u(x,t) $ enforces \Eq{14}, as in $\int_{-\infty}^{\infty} \frac{\rmd k}{2\pi} \rme^{i k x}=\delta(x)$; the final integral over $u(x,t+\delta t)$ insures that the latter takes  its appropriate value, proving \Eq{15}. 

Note that we have used the so-called It\^o discretization, where the r.h.s.\ of \Eq{14} is evaluated at time $t$. If one were to take $t+\delta t/2$ or $t+\delta t$, a non-trivial factor would appear. 

The strategy forward is now clear: define the path-integral measure 
\be
\int \ca D[\tilde u] \ca D[u] := \prod_x \prod_t  \frac{\eta}{\delta t} \int\limits_{-i \infty}^{i \infty} \frac{\rmd \tilde u(x,t)}{2\pi i}\int\limits_{-\infty}^\infty \rmd u(x,t+\delta t)
\ee 
and action  
\bea\label{S[u,utilde,F]}
\!\!\!{\cal S}[u,\tilde u,F] = \!\int_{x,t} \!\tilde u(x,t) \Big[& \big(\eta \partial_{t}  - \nabla^{2}+m^{2}\big) \big( u (x,t)-w\big) \nn\\
 &-F \big(x,u (x,t)\big)  \Big] .
\eea
The proper definition of $\eta \partial_t$ is as given in \Eq{15}. 
While $\tilde u(x,t)$ in the path-integral is purely imaginary, it can be continued analytically, as long as the 
integration path remains convergent.

The expectation of an observable  $\ca O$ (which can now depend on any of the variables in the measure) reads
\be
\left< \ca O \right>  = \int \ca D[\tilde u] \ca D[u] \,  \rme^{-\ca S[u,\tilde u,F]}\,\ca O.
\ee
The final step is to average over disorder, which by assumption is Gaussian with variance given in \Eq{Delta-def}. 
Denoting this by an overline, we obtain the disorder-averaged action 
$
\rme^{-{\cal S}[u,\tilde u]}  := \overline {\rme^{-S[u,\tilde u,F]}}
$, with 
\bea\label{dyn-action}
{\cal S}[u,\tilde u]& =& \int_{x,t} \!\tilde u (x,t)  (\eta\partial_{t}-\nabla^{2}+m^{2}) \big[ u
(x,t)-w\big] \nn\\
  &-&
\frac12 \int_{x,t,t'} \tilde u (x,t)\Delta_0 \big(u (x,t){-}u (x,t')\big)\tilde u
(x,t').\nn\\
\eea
For simplicity of notations, we put $\eta\to 1$ in the remainder of this work.
The response function is  defined as the answer of the system to a perturbation by the force $f(x,t)$, which we add to the r.h.s. of \Eq{EOM}, 
\bea\label{170}
R(x',t'|x,t) := \frac{\delta}{\delta f(x,t)} {\overline{u(x',t')} } = \left<   u(x',t') \tilde u(x,t)  \right>  . \nn\\
\eea 
While the overbar indicates a disorder average, the angular brackets denote averages w.r.t.\ the action \eq{S[u,utilde,F]}.
In a translationally invariant system, $R(x',t'|x,t) $ does only depend on $x'-x$ and $t'-t$, and is denoted  by
\be
R(x'-x,t'-t) := R(x',t'|x,t)  .
\ee
The most convenient representation  is the spatial  Fourier transform. For the  free theory it reads
\be\label{R0}
R(k,t) = \left< {u(k,t{+}t') \tilde u(-k,t')} \right> = \rme^{-(k^2+m^2)t} \Theta(t). 
\ee
This form allows us to   integrate over time, even in presence of a non-trivial time-behavior, as we will see to arise in the next section.

\subsection{Complications due to the non-analyticity of the disorder}
To perform the calculations, we define a graphical notation for the 
disorder vertex, 
\be\label{19}
{\parbox{.9cm}{{\begin{tikzpicture}
\coordinate (x1t1) at  (0,0) ; 
\coordinate (x1t2) at  (0,.5) ; 
\node (t3) at  (0,-.2)    {$\scriptstyle t$};
\node (t1) at  (0,.7)     {$\scriptstyle t'$};
\fill (x1t1) circle (2pt);
\fill (x1t2) circle (2pt);
\draw [dashed,thick] (x1t1) -- (x1t2);
\draw [enddirected]  (x1t1)--(0.5,0);
\draw [enddirected]  (x1t2)--(0.5,0.5);
\end{tikzpicture}}}} = \Delta\big(u(x,t)-u(x,t')\big) \tilde u(x,t) \tilde u(x,t').
\ee
The arrows represent the response fields $\tilde u(x,t) \tilde u(x,t')$, the dashed line the disorder
$\Delta\big(u(x,t)-u(x,t')\big) $, and integration over $x$, $t$ and $t'$ is implicit. 
(The spatial coordinate $x$ is not written.)

Let us illustrate the problem with one of the many 2-loop diagrams: 
\be\label{20}
{\parbox{2.cm}{{\begin{tikzpicture}
\coordinate    (x1t1) at  (0,0) ; 
\coordinate (x1t2) at  (0,.5) ; 
\coordinate (x2t3) at  (1,0) ; 
\coordinate (x2t4) at  (1,0.5) ; 
\coordinate (x3t5) at  (.25,1.2) ; 
\coordinate (x3t6) at  (.75,1.2) ; 
\node (t1) at  (-0.2,0)    {$\scriptstyle t_1$};
\node (t3) at  (1,-.2)    {$\scriptstyle t_3$};
\node (t5) at  (0.2,1.4) {$\scriptstyle t_5$};
\node (t6) at  (.8,1.4)    {$\scriptstyle t_6$};
\node (t2) at  (-.2,.5)    {$\scriptstyle t_2$};
\node (t4) at  (1,.7)   {$\scriptstyle t_4$};
\node (kpp) at  (0.5,.5)    {$\scriptstyle k+p$};
\node (mk) at  (0.45,-.2)   {$\scriptstyle k$};
\node (p) at  (0.03,.95)    {$\scriptstyle p$};
\node (p2) at  (.73,.95)    {$\scriptstyle p$};
\fill  (x1t1) circle (2pt);
\fill  (x1t2) circle (2pt);
\fill  (x2t3) circle (2pt);
\fill (x2t4) circle (2pt);
\fill (x3t5) circle (2pt);
\fill (x3t6) circle (2pt);
\draw [directed] (x1t1) -- (x2t3);
\draw [directed] (x1t2) -- (x3t5) ;
\draw [directed] (x2t3) -- (x1t2);
\draw [directed] (x2t4) -- (x3t5) ;
\draw [dashed,thick] (x1t1) -- (x1t2);
\draw [dashed,thick] (x2t3) -- (x2t4);
\draw [dashed,thick] (x3t5) -- (x3t6);
\draw [enddirected]  (x3t6)--(1.2,1.2);
\draw [enddirected]  (x2t4)--(1.4,0.5);
\end{tikzpicture}}}}
\ee
An arrow between two points represents the response function \eq{R0}, with the momentum it carries indicated, and time advancing in the direction of the arrow. 
An arrow entering into a vertex corresponds to a Wick contraction, and yields a derivative. Labeling the space coordinates at the bottom left  by $x$, 
bottom right by $y$, and top by $z$,  the diagram reads (up to a global prefactor) 
\bea
&&\int_{k,p}\int_{t_1}\int_{t_2}\int_{t_3}\int_{t_4} \Delta'\big(u(x,t_2)-u(x,t_1)\big) \nn\\
&& \times \Delta'\big(u(y,t_4)-u(y,t_3)\big) \Delta''\big(u(z,t_5)-u(z,t_6)\big) \qquad\nn\\
 &&  \times   R(k,t_3-t_1) R(k+p,t_2-t_3) R(p,t_5-t_2) R(p,t_5-t_4).\nn\\
\eea
Since the  response functions decay exponentially fast, they imply that $t_1\approx t_2\approx t_3\approx t_4 \approx t_5$, whereas $t_6$ and thus $u(z,t_5)-u(z,t_6)$ are   arbitrary time and arbitrary position differences respectively. Denoting this possibly large difference in position by $w$,   $\Delta''\big(u(z,t_5)-u(z,t_6)\big)\to \Delta''(w)$ can take any (allowed) value. This is not the case for the other arguments, e.g.\ 
\bea
\Delta'\big(u(x,t_2)-u(x,t_1)\big) &\to&  \Delta'(u) \mbox{~for $u>0$ small~}\nn\\
&\simeq& \Delta'(0^+) ,
\eea
since $t_2>t_1$ due to the causality of the response functions $R$, and $u(x,t)$ increases monotonically with time 
(Middleton theorem \cite{Middleton1992}, see section 3.3 of \cite{Wiese2021}).
The delicate factor is 
\be
\Delta'\big(u(y,t_4)-u(y,t_3)\big)\simeq \Delta'(0^+) \mbox{sign}(t_4-t_3),
\ee%
\begin{figure}[t]
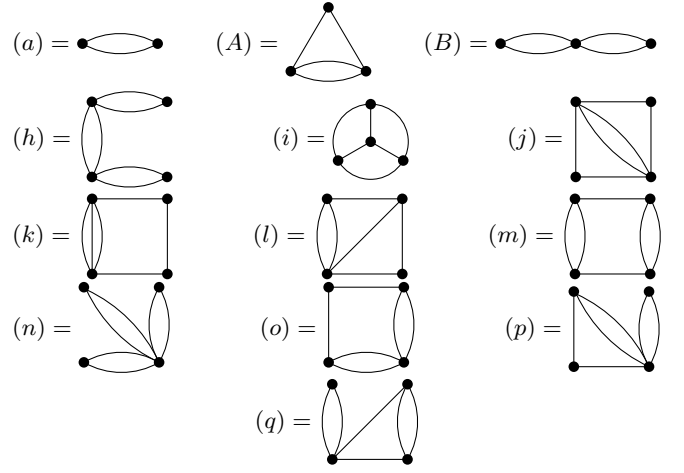

$(a)=\Ione$ \hfill $(A)=\ILH$  \hfill $(B)=\Ibanana$~

$(h)=\Ih \hfill (i)=\Ii \hfill (j)=\Ij~$

$(k)=\Ik \hfill (l)=\Il \hfill (m)=\IIm$

$(n)=\In \hfill (o)=\Io \hfill (p)=\Ip$

$(q)=\Iq $

\caption{Diagrams at 3-loop order (without insertion of lower order counter-terms)}
\end{figure}%
where we have again expanded for small times. 
It can have both signs. 
The integral to be performed is proportional to 
\bea\label{24}
&&\int\limits_{k,p}\int\limits_{-\infty}^{t_5}\!\!\rmd t_2
\int\limits_{-\infty}^{t_2}\!\!\rmd t_3  
\int\limits_{-\infty}^{t_3}\!\!\rmd t_1  
\int\limits_{-\infty}^{t_5}\!\!\rmd t_4 \,\rme^{-[m^2{+}k^2](t_3{-}t_1)} \nn\\
&&\times \rme^{-[m^2{+}(k{+}p)^2] (t_2{-}t_3)} \rme^{ -[m^2+ p^2](t_5-t_2)}\nn\\
&& \times  \rme^{- [m^2+p^2](t_5-t_4)}\mbox{sign}(t_4-t_3).
\eea
Without a sign function, all integrals can be evaluated via a single rule 
\be
\int_{-\infty}^t \rmd t'\, \rme^{-[m^2+k^2](t-t')} = \frac1{m^2+k^2}.
\ee
In contrast, \Eq{24} leads to the unusual combination
\bea\label{second-line}
&&
\int\limits_{k,p}\frac1{[k^2+m^2][(k+p)^2+m^2]^2[p^2+m^2]} \nn\\
&& -\int\limits_{k,p}\frac1{[k^2+m^2][(k+p)^2+m^2]^2[(k+p)^2+p^2+2 m^2]}.\nn\\ 
\eea
The first line is a standard diagram
\be\label{27}
\ILH=\int\limits_{k,p}\frac1{[k^2+m^2][(k+p)^2+m^2]^2[p^2+m^2]}.
\ee
In contrast, 
the second  line of \Eq{second-line} is a genuinely new contribution. 
What we will see in the following is that up to 3-loop order, for the effective disorder correlator, and the force at depinning,
all these new diagrams cancel. In contrast, in dynamic diagrams, i.e.\ those correcting the dynamic exponent $z$, these novel contributions appear.

In the next section, we list all diagrams contributing to the renormalization of $\Delta(w)$ up to 3-loop order. 
Each dynamic diagram, as plotted in \Eq{20}, reduces to a static  (momentum) diagram upon dropping the temporal information, i.e.\ dropping the direction in which an arrow goes, as well as the times at the vertex. 
Graphically this amounts to the {\em temporal reduction}
\be
{\parbox{2.cm}{{\begin{tikzpicture}
\coordinate    (x1t1) at  (0,0) ; 
\coordinate (x1t2) at  (0,.5) ; 
\coordinate (x2t3) at  (1,0) ; 
\coordinate (x2t4) at  (1,0.5) ; 
\coordinate (x3t5) at  (.25,1.2) ; 
\coordinate (x3t6) at  (.75,1.2) ; 
\node (t1) at  (-0.2,0)    {$\scriptstyle t_1$};
\node (t3) at  (1,-.2)    {$\scriptstyle t_3$};
\node (t5) at  (0.2,1.4) {$\scriptstyle t_5$};
\node (t6) at  (.8,1.4)    {$\scriptstyle t_6$};
\node (t2) at  (-.2,.5)    {$\scriptstyle t_2$};
\node (t4) at  (1,.7)   {$\scriptstyle t_4$};
\node (kpp) at  (0.5,.5)    {$\scriptstyle k+p$};
\node (mk) at  (0.45,-.2)   {$\scriptstyle k$};
\node (p) at  (0.03,.95)    {$\scriptstyle p$};
\node (p2) at  (.73,.95)    {$\scriptstyle p$};
\fill  (x1t1) circle (2pt);
\fill  (x1t2) circle (2pt);
\fill  (x2t3) circle (2pt);
\fill (x2t4) circle (2pt);
\fill (x3t5) circle (2pt);
\fill (x3t6) circle (2pt);
\draw [directed] (x1t1) -- (x2t3);
\draw [directed] (x1t2) -- (x3t5) ;
\draw [directed] (x2t3) -- (x1t2);
\draw [directed] (x2t4) -- (x3t5) ;
\draw [dashed,thick] (x1t1) -- (x1t2);
\draw [dashed,thick] (x2t3) -- (x2t4);
\draw [dashed,thick] (x3t5) -- (x3t6);
\draw [enddirected]  (x3t6)--(1.2,1.2);
\draw [enddirected]  (x2t4)--(1.4,0.5);
\end{tikzpicture}}}}
~~\stackrel{\text{temporal reduction}} 
{-\!\!\!\!-\!\!\!-\!\!\!-\!\!\!-\!\!\!-\!\!\!-\!\!\!-\!\!\!\longrightarrow}~~~ \ILH.
\ee
In order to alleviate the notations, we only draw  the {\em temporally reduced} (static) representation for each diagram in the next section. This should not be confounded with the momentum integral itself. 
Surprisingly, for each correction to the disorder, the only momentum integral which survives after summation over all temporal configurations is the {\em temporally reduced diagram read as a momentum integral}, as in \Eq{27}.

\begin{figure*}
\begin{eqnarray*}
\Ione: -\!\!\!\int\!\!\!\!\!\int_w   (a) &=&  \frac12 \Big[\Delta(0)-\Delta(w)\Big]^2 I_1\\
\ILH:-\!\!\!\int\!\!\!\!\!\int_w   (A) &=&  \Big[ \Delta '(0^+)^2 \Delta (w)\black + 
   \Big(\Delta (0)-\Delta (w)\Big) \Delta
   '(w)^2
 ~ \underline{-  2 \Delta '(0^+)^2 \Delta(w)} \Big]   I_A
\\
\Ibanana: -\!\!\!\int\!\!\!\!\!\int_w   (B) &=&  \Big[ \underline{\Delta '(0^+)^2 \Delta (w)} -\frac{1}{2} \Big(\Delta
   (0){-}\Delta (w)\Big)^2 \Delta ''(w)\Big]   I_B  
\\
\Ih: -\!\!\!\int\!\!\!\!\!\int_w   (h) &=&  \Big[ \frac{1}{2} \Big(\Delta (w){-}\Delta (0)\Big)^2
   \Delta ''(w)^2 
   ~ \underline{- \Delta'(0^+)^2 \int\!\!\!\!\!\int_w  \Delta''(w)^2 } \Big] I_h
\\
\Ii: -\!\!\!\int\!\!\!\!\!\int_w   (i) &=&  \Big[ \frac{1}{2} \Delta '(w)^4-\Delta '(0^+)^2 \Delta '(w)^2 
 ~  \underline{+4 \Delta'(0^+)^2 \Delta ''(0) \Delta (w)} \Big] I_i
\\
\Ij: -\!\!\!\int\!\!\!\!\!\int_w   (j) &=&  \Big[   \Big(\Delta (w){-}\Delta (0)\Big)^2
   \Delta ''(w)^2 
   ~ \underline{-2 \Delta'(0^+)^2 \int\!\!\!\!\!\int_w  \Delta''(w)^2 } \Big] I_j\\
\Ik: -\!\!\!\int\!\!\!\!\!\int_w  (k)&=& 0\\
\Il: -\!\!\!\int\!\!\!\!\!\int_w   (l) &=& 4 \Big[ \Delta '(w)^2 \big( \Delta (w){-}\Delta (0)\big)
   \Delta ''(w) 
   - \Delta '(0^+)^2 \Delta ''(0) \Delta (w)
   \nn\\
   &&  ~~~ \underline{+  \Delta '(0^+)^2 \Delta ''(0) \Delta (w) + \Delta '(w)^2\Delta
   '(0^+)^2-\Delta '(0^+)^2 \int\!\!\!\!\!\int_w   \Delta
   ''(w)^2  }   \Big] I_l \\
\IIm: -\!\!\!\int\!\!\!\!\!\int_w   (m) &=& \Big[ \frac{1}{2} \Delta '(w)^4-\Delta
   '(0^+)^2 \Delta '(w)^2 
   \underline{+ \,  6 \Delta '(0^+)^2 \Delta ''(0) \Delta(w) 
   -\Delta '(0^+)^2 \int\!\!\!\!\!\int_w   \Delta
   ''(w)^2 }
   \Big] I_m \\
\In: -\!\!\!\int\!\!\!\!\!\int_w (n) &=& \Big[\frac{1}{6} \Delta ^{(4)}(w) \Big(\Delta(w)-\Delta (0)\Big)^3
\nn\\
&& 
\underline{ +\Delta '(0^+)^2
   \Big(\big(\Delta (0)-\Delta (w)\big) \Delta ''(w)+\Delta'(w)^2\Big) 
    -\Delta '(0^+)^2 \int\!\!\!\!\!\int_w   \Delta
   ''(w)^2  }\Big] I_n \\
\Io: -\!\!\!\int\!\!\!\!\!\int_w   (o) &=&  \Big[ \Delta '(w)^2 \big( \Delta (w){-}\Delta (0)\big)
   \Delta ''(w)  - \Delta '(0^+)^2 \Delta ''(0) \Delta (w)
   \nn\\
   &&  ~~ \underline{+  \Delta '(0^+)^2 \Delta ''(0) \Delta (w) + \Delta '(w)^2\Delta
   '(0^+)^2 -\Delta '(0^+)^2 \int\!\!\!\!\!\int_w   \Delta
   ''(w)^2  }   \Big] I_o  \\
\Ip: -\!\!\!\int\!\!\!\!\!\int_w   (p) &=&2 \Big[ \Big(\Delta (w)-\Delta (0)\Big)^2 \Delta
   ^{(3)}(w) \Delta '(w) \nn\\
&& ~~+ \underline{  \Delta '(0^+)^2 \Big(\Delta (w)-\Delta (0)\Big) \Delta ''(w) 
 -2 \Delta '(0^+)^2 \Delta '(w)^2    + 3 \Delta '(0^+)^2 \int\!\!\!\!\!\int_w 
   \Delta ''(w)^2}
  \black \Big]I_p \\
\!\!\!\Iq: -\!\!\!\int\!\!\!\!\!\int_w   (q) &=& 
\Big[ \big(\Delta(w)-\Delta(0)\big) \Delta''(w) \big( \Delta'(w)^2{-} \Delta'(0^+)^2
\big) \underline{ -   6 \Delta '(0^+)^2 \Delta ''(0) \Delta(w)  +2\Delta '(0^+)^2 \int\!\!\!\!\!\int_w   \Delta
   ''(w)^2 } \Big]I_{q} \nn\\
 \end{eqnarray*}
\caption{All diagrams correcting the disorder up to 3-loop order. All $\Delta$ are bare $\Delta_0$, with the index suppressed for compactness of notation. While we calculated the corrections to $\Delta(w)$, we report its integrated form $\delta R(w):=-\int_0^w \rmd w' \int_0^{w'}\rmd w''\, \delta \Delta(w)$ for compactness. This is the correction to the potential correlator $R(w)$. The non-underlined terms are present in the statics \cite{WieseHusemannLeDoussal2018}, the underlined ones are additional contributions at depinning.
We note that the following expressions are proportional to each other, 
$(o)\sim (l)$, and $(h) \sim (j)$. The momentum integrals, which correspond to the icons in the same line, are given in appendix \ref{app:Integrals}.
}
\label{disorder-table}
 \end{figure*}

\subsection{Diagrams correcting the disorder}
Denoting by $\delta^{(\ell)}$ the contributions at $\ell$-loop order, 
the corrections to the disorder up to 3-loop order are given by 
\bea
\delta^{(1)} \Delta(w) &=&  (a), \\
\delta^{(2)} \Delta(w) &=& (A) + (B),\\
\delta^{(3)} \Delta (w) &=& (h)+ (i)+ (j)+ (k)+ (l)+ (m)+ (n)\nn \\
&& + (o)+ (p)+ (q).
\eea
The different diagrams are given on Fig.~\ref{disorder-table}.
To simplify the expressions and for easier comparison with the statics, we write the diagrams as minus a total second derivative, s.t.\ the expression would be the correction to the potential correlator $R(w)$ (i.e.\ $\Delta(w) = - R''(w)$). 
The additional terms at depinning, as compared to the statics, are underlined. 
We note that not all terms can be integrated explicitly, the notable exception being $\sim  \Delta ''(w)^2 $.

\section{The $\beta$-function and its fixed point}\label{The beta-function and its fixed point}
\subsection{The $\beta$-function}
Using the above diagrams and the  integrals tabulated in appendix \ref{app:Integrals}, we   write   the dimensionfull effective disorder correlator $ \Delta_{\rm eff}(w)$
\be\label{18}
\Delta_{\rm eff}(w) = \Delta_0(w) + 
\delta^{(1)} \Delta(w) + 
\delta^{(2)} \Delta(w) +
\delta^{(3)} \Delta (w) + ...
\ee
The r.h.s.\ is a function of the bare disorder $\Delta_0(w)$, its derivatives,  and $m$. 
The $\beta$-function for the renormalized (effective) dimensionfull disorder correlator $\Delta_{\rm eff}(w) $ as a function of the bare disorder $\Delta_0(w)$ is defined as 
\bea
\partial_\ell\Delta_{\rm eff}(w) &:=& -m \partial_m \Delta_{\rm eff}(w) \big|_{\Delta_0} \\
&=&\epsilon\left[   
\delta^{(1)} \Delta(w) + 2\delta^{(2)} \Delta(w) + 3\delta^{(3)} \Delta (w) +...\right] \nn
\eea
There are two steps left: first,  rewrite \Eq{18} as a rule 
\be\label{Delta0-to-Delta-eff}
 \Delta_0(w) \to \Delta_{\rm eff}(w) -  
\delta^{(1)} \Delta(w) - 
\delta^{(2)} \Delta(w) -
\delta^{(3)} \Delta (w) , 
\ee
where  as above $\delta^{(i)}\Delta(w)$ are functions of the bare disorder $\Delta_0(w)$ (and its derivatives). Applying this rule three times to $\partial_\ell \Delta_{\rm eff}(w)$ 
gives $\partial_\ell \Delta_{\rm eff}(w)$  as a function of $\Delta_{\rm eff}(w)$ instead of $ \Delta_0(w)$.  

In a second step,   define 
\be\label{21}
\tilde \Delta(w) := \epsilon I_1 m^{2\zeta}  \Delta_{\rm eff}(w m^{-\zeta}).
\ee
This rescaling with $\epsilon I_1$ and the roughness exponent $\zeta$ allows to obtain a fixed point. 
Rescaling with $\epsilon I_1\sim m^{-\epsilon}$ instead of $m^{-\epsilon}$  eliminates cumbersome numerical factors. 

This yields the $\beta$-function for the renormalized dimensionless disorder $\tilde \Delta(w)$, 
\begin{widetext}
\bea
\label{beta-3-loop}
\partial_\ell \tilde \Delta(w) &=& (\epsilon - 2\zeta) \tilde \Delta(w) + \zeta w \tilde \Delta'(w) -\partial_w^2
\Big[ \frac{1}{2} (\tilde \Delta (0)-\tilde \Delta (w))^2\Big] \nn\\
&&- \partial_w^2 \Big[ \Big(-  \frac12  -\frac{\epsilon }{4}+C_3 \epsilon \Big)  \Big(\tilde \Delta '(0^+)^2 \tilde \Delta
   (w)+(\tilde \Delta (w)-\tilde \Delta (0)) \tilde \Delta
   '(w)^2\Big)\Big] \nn\\
&& - \partial_w^2\Big[ \frac{3}{4} \zeta (3) \Big(\tilde \Delta '(w)^4-2
   \tilde \Delta '(0^+)^2 \tilde \Delta '(w)^2+8 \tilde \Delta '(0^+)^2
   \tilde \Delta ''(0) \tilde \Delta (w)\Big) \nn\\
&&\qquad ~~ + 2 \tilde \Delta '(w)^2 \Big(\tilde \Delta '(0^+)^2+(\tilde \Delta
   (w)-\tilde \Delta (0)) \tilde \Delta ''(w)\Big)   \nn\\
&&\qquad ~~ + C_3 \Big((\tilde \Delta (w)-\tilde \Delta (0))^2 \tilde \Delta
   ''(w)^2-\frac{1}{2} \tilde \Delta '(w)^4+(\tilde \Delta
   (0)-\tilde \Delta (w)) \tilde \Delta '(w)^2 \tilde \Delta ''(w)-6
   \tilde \Delta '(0^+)^2 \tilde \Delta ''(0) \tilde \Delta (w)\Big)   
   \Big] \nn\\
&&+2  \tilde \Delta '(0^+)^2     \tilde \Delta ''(w)^2  + \ca O(\tilde \Delta^5) 
\eea\end{widetext}
The amplitude $C_3$ involves   $\psi(x) := \partial_x \ln \Gamma(x)$, 
\be
C_3 = \frac{\psi'(\frac13)}6-\frac{\pi^2}9.
\ee
The first two terms are a consequence of the rescaling \eq{21}, while the remaining ones 
are the direct loop corrections: the 1-loop term is on the first line, the 2-loop terms on the second line, followed by the 3-loop contributions. 

Let us compare this $\beta$ function  to the 3-loop result in   equilibrium, obtained in Ref.~\cite{WieseHusemannLeDoussal2018}.
 We   see that the flow equations differ by anomalies terms (terms proportional to $\Delta'(0^{+})^2$). All additional terms (at depinning, as compared to equilibrium) are underlined  in Fig.~\ref{disorder-table}. Let us already point out that in equilibrium the RF-fixed point discussed in the next section has a trivial exponent of $\zeta=\epsilon/3$ to all orders.
\subsection{Fixed point} 
\Eq{beta-3-loop} has a discrete set of fixed points, among which   one is fully attractive, and represents the dominant random-field universality class, see e.g.\  \cite{Wiese2021}. While we could in principle follow the flow to this attractive fixed point, it is better to directly write down the fixed-point equation, which gives $\zeta$ and $\tilde \Delta(w)$ to 3-loop order. While at 1-loop order we can do this analytically, 
much of the information for 2-loop and 3-loop order has to be obtained numerically. Useful analytic constraints are obtained by integrating \Eq{beta-3-loop} over $w$, 
\bea\label{integrated-beta-function}
0 &=& \int_0^\infty \partial_\ell \tilde \Delta(w) \rmd w\nn\\
&=&  (\epsilon- 3 \zeta) \int_0^\infty \tilde \Delta(w) \rmd w \nn\\
&&-   \left(1 -2 C_3 \epsilon +\frac{\epsilon }{2} \right) \tilde \Delta '(0^+)^3 \nn\\
&&+ 3 \Big(2-3 C_3+2 \zeta (3) \Big) \tilde \Delta'(0^+)^3 \tilde \Delta ''(0)   \nn\\
&&+ 2  \tilde \Delta '(0^+)^2  \int_0^\infty   \tilde \Delta ''(w)^2  \rmd w. 
\eea
We used that $\tilde \Delta(w)$ is decaying fast to zero for $w\to \infty$, thus all boundary terms at infinity vanish. In particular $\lim_{w\to \infty} w \tilde \Delta(w)=\lim_{w\to \infty} \tilde \Delta'(w) = ... = 0$.
We make the ansatz
\bea\label{Delta-ansatz}
\tilde \Delta(w) &=& \frac{\epsilon}3 y(w) + \frac{\epsilon^2}{18}  y_2(w) + \epsilon^3   y_3(w) \\
\zeta &=& \frac{\epsilon}3 + \zeta_2 \epsilon^2 + \zeta_3 \epsilon^3 + ... \\
y(0) &=& 1, \quad y_2(0)=0, \quad y_3(0)=0. 
\eea
(The numerical factors $1/3$ and $1/18$ are for historical reasons, to agree with the 
conventions of  \PLDKW).

\subsection{1-loop order}
After integrating the 1-loop solution twice, this yields (see e.g.\ \cite{Wiese2021}, section 2.6)
\be
\frac{w^2}{2}-y(w)+\log (y(w))+1 = 0. 
\ee
This is a simple expression for $w(y)$. 
Mathematica knows the inverse function $y(w)$ as a ProductLog, 
\be
y(w) = -W\left(-\rme^{-\frac{w^2}{2}-1}\right).
\ee
Its series expansion is 
\bea\label{y-series}
y(w) &=& 1-w+\frac{w^2}{3}-\frac{w^3}{36}-\frac{w^4}{270
   }-\frac{w^5}{4320}+\frac{w^6}{17010}\nn\\
&&+\frac{139w^7}{5443200} +\frac{w^8}{204120}+\frac{571
   w^9}{2351462400} \nn\\
&&-\frac{281
   w^{10}}{1515591000}+ ... 
\eea
Integrals we need later are 
\bea
a_1&:=&\int_0^\infty \rmd w \, y(w) = 0.7753042451883378, \\
a_2&:=&\int_0^\infty \rmd w \,  y''(w)^2 = 0.44750763980522135.\quad 
\eea
Simple analytical integral representations are obtained by converting the $w$ integrals into $y$ integrals:
\bea
a_1 &=& \sqrt{2}  \int_0^1 \sqrt{y-\log (y)-1} \, \rmd y, \\
a_2 &=& \frac23 +\sqrt{2} \int_0^1 \frac{ y \sqrt{y-\log (y)-1} }{(y-1)^5}  \nn\\
&& \qquad ~~~~~\times \Big[(y-1) (y+5)-2 (2 y+1) \log (y)\Big]\, \rmd y. \nn\\
\eea

\subsection{2-loop order}
As a first consequence of the integral relation \eq{integrated-beta-function}  we find to order $\epsilon^2$  
\be
-   \zeta_2 \int_0^\infty  {y(w)} \,\rmd w - \left( \frac{y'(0)}{3} \right)^3=0
\ee
\Eq{integrated-beta-function} at 2-loop order then yields
\bea \zeta_2 &=& \frac1{27 a_1}\nn\\
& =& 0.04777097154682305779461454163450931593852 \nn\\
&=& \frac{0.1433129146404691733838436249035279478156}3 .\nn\\
\eea
We then need $y_2(w)$. A good approximation is obtained by solving the 2-loop $\beta$-function perturbatively around 0, and then producing a fit for $y_2(w)/y(w)^2$, 
\bea\label{y2-approx}
y_2(w) &\approx& \Big(-1.14012 w - 1.31245 w^2 - 0.927184 w^3 \nn\\
&& - 0.509678 w^4 - 
 0.23776 w^5 - 0.0983357 w^6  \nn\\
 && - 0.0370205 w^7 - 0.0129135 w^8 - 
 0.00422806 w^9 \nn\\
 && - 0.00131226 w^{10}+ ...\Big) y(w)^2.
\eea 
A second approximation stems from the observation that $y_2(w) \approx \mbox{const}\, w y'(w)$, which would arise when the second-order solution just changes its amplitude, and this amplitude change is absorbed via a rescaling, sending $w\to w[1+ \ca O(\epsilon)]$. 
We can therefore write (with more terms used in practice)
\be\label{y2-approx-2}
y_2(w) \approx \big[1.14012  + \ca O(w) \big] w y'(w) . 
\ee
Another approximation is to do a Taylor expansion on $y_2(w)/(w y'(w))$, and then use the diagonal Pad\'e for its approximation. We show   for illustration a relative low-order approximant, 
\be\label{y2-approx-3}
y_2(w) = 
\frac{1.14012 {-}0.597926
   w  {+} 0.0931393 w^2+ ...}{1{-}0.342252 w{+} 0.0465221 w^2+...}\, w y'(w).
\ee
Later we need
\be
a_3:=\int_0^\infty y_2(w) \, \rmd w  =-0.636336 (1\pm 7 \times 10^{-5}).
\ee
The error bar is from a numerical solution of the FP equation, combined with the 
approximations \eq{y2-approx}--\eq{y2-approx-3}. 

\subsection{3-loop order}
The   integral relation \eq{integrated-beta-function}  to next order reads
\bea
0&=& - \zeta_3 a_1- \frac{\zeta_2}6 a_3  \nn\\
&&- \left(-2 C_3 +\frac12 \right)\frac{y'(0)^3}{3^3} - 3 \frac{y'(0)^2}{3^2}
\frac{y_2'(0)}{ 18} \nn\\
&&+ 3 \Big(2-3 C_3+2 \zeta (3) \Big) \frac{y'(0)^3}{3^3} \frac{y ''(0)}{ 3}   \nn\\
&&+ 2  \frac{y '(0)^2}{3^2}  \frac{a_2}{3^2} .
\eea
Inserting everything we calculated above, we find\footnote{Note that these relation are rather sensitive to muddling with coefficients: multiplying any of the coefficients with a factor of $b$ shows that the solution depends on $b$ with a factor of at least $0.1b$ and max $1.3b$. So an error made in $a_1$, $a_2$ and $a_3$ has a strong impact on the final result.}
\bea
\label{zeta-3-values}
\zeta_3 = \left\{ 
\begin{array}{cl}
-0.0683545 & \mbox{~from \Eq{y2-approx} at order 30}
\\
-0.0683547 & \mbox{~from \Eq{y2-approx} at orders 30 to 40~~~~~~}
\\
-0.068354436 & \mbox{~from \Eq{y2-approx-2} at order 30}
\\
-0.068354414 & \mbox{~from \Eq{y2-approx-3} from $\mbox{Pad\'e}_{15,15}$}
\\
-0.0683544 & \mbox{~with $a_3$ from shooting for $y_2(w)$}
\end{array}~~~~
\right. 
\eea
Solving the $\beta$-function numerically via shooting,  we find  
\be
\zeta_3 = -0.0683544.
\ee
This value is especially consistent with the last value in \Eq{zeta-3-values}.
The relative difference of the above values is better than $10^{-5}$.

We can also create a series expansion for $y_3(w)$, as we did for $y_2(w)$. 
Doing this and using shooting with $\Delta(w=4)=0$  (instead of $\Delta(\infty)=0$), we find (probably less reliable)
\be
\zeta_3 = -0.0683803.
\ee
Neglecting this last value, our confidence for $\zeta_3$ is 
\be
\zeta_3 = -0.0683544 (2).
\ee

\subsection{Numerical values and resummation}
\begin{figure}[t]
\Fig{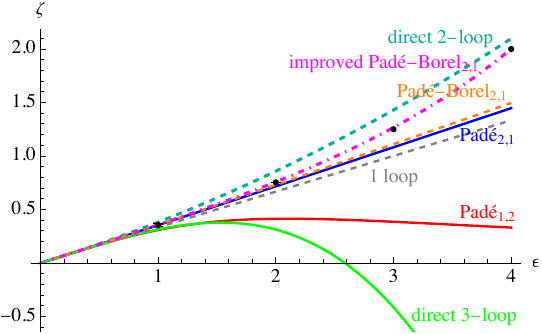}
\caption{$\zeta(\epsilon)$ in different schemes: 1-loop (black, dashed), direct 2-loop (cyan), direct 3-loop (green), as well as two Pad\'e appoximants, $\mbox{Pad\'e}_{1,2}$ (red) and $\mbox{Pad\'e}_{2,1}$ (blue). For the latter, we also show the Pad\'e-Borel resummation as explained in the main text.
The black dots are the result of numerical simulations for $d=2,3$, and the exact values $\zeta=5/4$ in $d=1$ as well as $\zeta=2$ in $d=0$. 
}
\label{f:zetaofepsilon}
\end{figure}
Fig.~\ref{f:zetaofepsilon} shows both the direct sum, as well as various approximations.  
\begin{table}[b]
\begin{tabular}{|c|c|c|c|c|}
\hline
method & $d=0$ & $d=1$  & $d=2$ & $d=3$
\\
\hline\hline
numeric/exact & 2  & 5/4   & $ 0.753(2)$   & $0.355(10)$ \\\hline
1-loop &  4/3 & 1   &  2/3  & 1/3  \\\hline
direct 2-loop & $2.09767$ & $1.42994$ & $0.85775$ & $0.38110$  \\\hline
direct 3-loop &  $-2.27702$ &  $-0.41563$ & $0.31092$ & $0.31275$ \\\hline
$\mbox{Pad\'e}_{1,2}$ & 0.33033 & 0.38454 & $0.41260$ & $0.30799 $\\\hline
$\mbox{Pad\'e}_{2,1}$ & 1.44701 & 1.08123 & $0.71615$ & $0.35299$ \\\hline
$\mbox{Pad\'e-Borel}_{2,1}$ & $1.47806$ & $1.10053$ & $0.72539$ & $0.35512$  \\ \hline
improved & $2$ & $1.26567$ & $0.75341$ & $0.35716$  \\
\hline
twice improved & $2$ & $1.25$ & $0.75182$ & $0.35658$  \\
\hline
\end{tabular}
\caption{Comparison of the various approximations for $\zeta$. The simulation data in dimensions $d=2$ and $d=3$ are from  \cite{RossoHartmannKrauth2002}. Results in $d=0$ from \cite{LeDoussalWiese2008a,terBurgWiese2020}, $d=1$ from \cite{GrassbergerDharMohanty2016,ShapiraWiese2023}.} 
\label{zeta-table}
\end{table}
From the various Pad\'e approximants, the best one is $\mbox{Pad\'e}_{2,1}$:
\be
\zeta(\epsilon) \approx \epsilon \frac{\frac{1
   }{3} + 0.524731 \epsilon }{1+ 1.43088 \epsilon} + \ca O(\epsilon^4). 
\ee
It is the only Pad\'e-approximant which is monotone for large $\epsilon$. 

A similar approximant can be used for a Pad\'e-Borel resummation. To this aim define 
\be
\zeta^{\rm Borel}(t):= \sum_{n=1}^\infty \frac{\zeta_n t^n}{n!}
\ee 
This series stops at order $t^3$ (3-loop order). 
As above, the Pad\'e-approximant which behaves well for large $t$ is ${\text{Pad\'e}_{2,1}}$
\be
\zeta^{\rm Borel}_{\text{Pad\'e}_{2,1}}(t)=\frac{\frac{t}{3}+ 0.182872 t^2}{1+0.47696 t}.
\ee
Using this, we obtain an approximation\footnote{The reader can verify that when expanded in $\epsilon$, this equation gives back the original series.} for $\zeta$, 
\be
\zeta^{\text{Pad\'e-Borel}}(\epsilon) := \int_0^\infty \frac{\rmd t}{\epsilon}\, \zeta^{\rm Borel}_{\text{Pad\'e}_{2,1}}(t) \,\rme^{-t/\epsilon} .
\ee
Let us use this as reference for the best 3-loop approximation. We remark that in $d=3$ there is a marked improvement, and the $\epsilon$-expansion result is now spot on the numerical solution, probably even more precise than the latter. 
In $d=2$ the improvement in precision is also noticable, with a relative deviation of less than $4\%$. In $d=1$ the relative error is now at  $12\%$, while $d=0$ is out of reach.

To improve the precision, we can use the information in $d=0$, where $\zeta=2$ (with $\sqrt{\ln}$ corrections); this fixes the coefficient of an additional quartic term, 
\be
\zeta^{\text{improved}}(\epsilon) = \zeta^{\text{Pad\'e-Borel}}(\epsilon)  + 0.00203884 \epsilon^4. \
\ee
With this correction, the prediction in $d=1$ becomes $1.266$, very close to the analytically known value of $\zeta=5/4$ \cite{ShapiraWiese2023}.
Using in addition   $\zeta(d=1)=5/4$, we find
\bea
\zeta^{\text{twice improved}}(\epsilon) &=& \zeta^{\text{Pad\'e-Borel}}(\epsilon)  + 0.00126488 \epsilon^4 \nn\\
&& + 0.00019349 \epsilon^5.
\eea
This is summarized in table \ref{zeta-table}.
Our best predictions and error estimates for the unknown dimensions $d=2$ and $d=3$ thus are
\bea
\zeta^{\rm best}_{d=2} &=& 0.752(1),  \\
\zeta^{\rm best}_{d=3} &=&0.357(1).
\eea

\subsection{The $\beta$-function in minimal subtraction}
The minimal subtraction scheme takes a prominent role in high-order RG calculations. How can this be implemented here?
The idea is to make an ansatz for $\Delta_{\rm r}(w)$ as a functional of $\Delta_0(w)$, and then to write the effective  $\Delta_{\rm eff}(w)$ in \Eq{18} 
as a function of $\Delta_{\rm r}(w)$, keeping only singular terms (minimal subtraction). Since  $\Delta_{\rm eff}(w)$ is an observable, it must be finite when expressed in terms of  $\Delta_{\rm r}(w)$.
This uniquely fixes $\Delta_{\rm r}[\Delta_0]$. 
Let us make the ansatz
\bea
&& \Delta_{\rm r}(w) =    \Delta_0(w) + 
\delta^{(1)} \Delta(w) + 
\ca S\circ \delta^{(2)} \Delta(w) +
\delta^{(3)} \Delta (w) \nn\\
&&  +  
\frac{1}{8} (\ca F\circ I_A) I_1 \Big[7
   \big(\Delta _0(w)-\Delta _0(0)\big) \Delta
   _0'(w){}^2 \Delta _0''(w) \nn\\
   && +\Delta _0'(0){}^2
   \Big(6 \Delta _0''(0) \Delta _0(w)+\big(\Delta
   _0(w)-\Delta _0(0)\big) \Delta
   _0''(w)\Big)\nn\\
   && +\Delta _0'(w){}^4+2 \big(\Delta
   _0(0)-\Delta _0(w)\big){}^2 \Delta
   _0{}^{(3)}(w) \Delta _0'(w)\Big]\nn\\
&&   + ...
\eea
Here $\ca S $ extracts the singular (in $\epsilon$) part of a diagram, while $\ca F$ extracts its finite part, $(\ca S + \ca F)\circ I \equiv I$.
The expression in the square brackets is what is obtained if one inserts the 1-loop expression into the 2-loop expression (repeated counter-term). 
This operation is successful, as  
\bea
&&\!\! \Delta_{\rm eff} (w)= \Delta_{\rm r}(w) {+} \frac{4 C_3{-}1}{8} \times \nn\\
&&\!\! \times \Big[2
   \big(\Delta_{\rm r}(0){-}\Delta_{\rm r}(w)\big) \Delta_{\rm r}'''(w) \Delta_{\rm r}'(w) \nn\\
   &&\!\! ~~~{-}\Delta_{\rm r}''(w)
   \big(\Delta_{\rm r}'(0){}^2{+}5 \Delta_{\rm r}'(w){}^2{+}2
   \big(\Delta_{\rm r}(w){-}\Delta_{\rm r}(0)\big) \Delta_{\rm r}''(w)\big)\Big].\nn\\
\eea
The ensuing  $\beta$-function is   longer than that in \Eq{beta-3-loop}, and we refrain from putting it here. 
It is more interesting to look at the difference, 
\bea
&&\partial_\ell \tilde \Delta(w) - \partial_\ell \Delta_{\rm r}(w)\Big|_{\Delta_{\rm r}=\tilde \Delta} \nn\\
&& = -\frac{4 C_3{-}1}{4}  \partial_w^2 \Big[     \epsilon
    \big(\tilde \Delta (0)-\tilde \Delta (w)\big) \big(\tilde \Delta
   '(0)^2+\tilde \Delta '(w)^2\big) \Big]  \nn\\
&&    -\frac{4 C_3{-}1}{8} \partial_w^2 \Big[    \tilde \Delta '(w)^4\nn\\
&&\qquad  +2 (\tilde \Delta (w)-\tilde \Delta (0)) \tilde \Delta
   '(w)^2 \tilde \Delta ''(w)\nn\\
&&\qquad +2 (\tilde \Delta (0)-\tilde \Delta (w))\times
   \nn\\
&&\qquad\times\Big((\tilde \Delta (w)-\tilde \Delta (0)) \tilde \Delta ''(w)^2-6
   \tilde \Delta'(0^+)^2 \tilde \Delta ''(0)\Big)  \Big] .\qquad 
\eea
A consistency check is  that this yields the same  $\zeta_3$. 
Integrating this equation over all $w$ yields
\bea
&& \int_0^\infty  \partial_\ell \tilde \Delta(w) - \partial_\ell \Delta_{\rm r}(w)\Big|_{\Delta_{\rm r}=\tilde \Delta} \rmd w \nn\\
&& = \frac{4 C_3{-}1}{2}  \epsilon
    \tilde \Delta'(0^+)^3+\frac{9 (1{-}4 C_3)}{4} 
    \tilde \Delta'(0^+)^3 \tilde \Delta ''(0).
\eea
Using \Eqs{Delta-ansatz} and \eq{y-series} shows that this vanishes at the required order $\epsilon ^4$. Thus $\zeta$ is independent of the scheme up to 3-loop order.

\section{The critical force}
\label{s:The critical force}

\newcommand{\tadpole}{\parbox{.3cm}{\fig{.3cm}{tadpole}}}
\newcommand{\sunset}{\parbox{3cm}{\fig{3cm}{sunset}}}
While renormalization of the disorder was already considered in the original 2-loop calculation \cite{ChauveLeDoussalWiese2000a,LeDoussalWieseChauve2002}, the dependence of the 
critical force at depinning was only considered in simulations \cite{RossoLeDoussalWiese2006a}, but not via RG. 
Here we address this issue. Since this calculation is novel, we give explicit results for each of  the 
dynamic diagrams involved up to 2-loop order.

\subsection{1 loop}
 The diagram in question is
\bea
\lefteqn{\parbox{1.1cm}{{\begin{tikzpicture}
\coordinate (x1t1) at  (0,0) ; 
\coordinate (x1t2) at  (0,1) ; 
\node (x) at  (0,0)    {$\!\!\!\parbox{0mm}{$\raisebox{-3.5mm}[2.5mm][2.5mm]{$\scriptstyle x$}$}$};
\node (t1) at  (-.25,0)    {$\!\!\!\parbox{0mm}{$\raisebox{-1mm}[0mm][0mm]{$\scriptstyle t_1$}$}$};
\node (t2) at  (-.25,1.2)    {$\!\!\!\parbox{0mm}{$\raisebox{-1mm}[0mm][0mm]{$\scriptstyle t_2$}$}$};
\fill (x1t1) circle (2pt);
\fill (x1t2) circle (2pt);
\draw [directed] (x1t1) arc(-90:90:0.5);
\draw [dashed,thick] (x1t1) -- (x1t2);
\draw [enddirected]  (x1t2)--(-.5,1);
\end{tikzpicture}}}} \nn\\
&&= \tilde u(x,t_2)\,\int_{t_1,k}\Delta_0'\big(u(x,t_2)-u(x,t_1)\big)\rme^{-(t_2-t_1)(k^2+m^2)}\nn\\
&&\hp{= \tilde u(x,t_2)\,\int_{t_1,k}}  \times \Theta(t_1<t_2)\nn\\
&&\simeq \tilde u(x,t_2)\,\int_{t_1,k}\Big[\Delta_0'(0^+)+\Delta_0''(0^+) (t_2{-}t_1)\dot u(x,t_2)\,{+} ...\Big]\nn\\
&& \hp{\simeq \tilde u(x,t_2)\,\int_{t_1,k}\Big[}\times\rme^{-(t_2-t_1)(k^2+m^2)}\Theta(t_1<t_2) \nn\\
&&= \tilde u(x,t_2)\,\int_{k}\frac{ \Delta_0'(0^+) }{k^2+m^2}+\frac{\Delta_0''(0^+)}{(k^2+m^2)^2}  \dot u(x,t_2)+ ...
\label{7.32}
\eea
The first term is the correction to the critical force, the second term the correction to 
friction. 
In summary,  
\bea
\delta f_{\rm c}^{(1)} &=& \Delta'(0^+) I_{\rm tp}, \\
I_{\rm tp} &=& \ITP = \int_k \frac{1}{k^2+m^2} = \frac{2
   m^{2}}{(d-4)
   (d-2)} (\epsilon I_1).\qquad 
\eea
See appendix \ref{s:Itp} for the integral.

\subsection{2 loop}
At 2-loop order, there a seven contributions to the critical force. Including all combinatorial factors, these read
\bea
\delta^{(2)}f_{\rm c} &=& \ca F_1+\ca F_2+\ca F_3+\ca F_4+\ca F_5+\ca F_6+\ca F_7\\
\ca F_1 &=&  {\parbox{2cm}{{\begin{tikzpicture}
\coordinate (x1t1) at  (0,0) ; 
\coordinate (x1t2) at  (0,.75) ; 
\coordinate (x2t3) at  (1.25,0) ; 
\coordinate (x2t4) at  (1.25,0.75) ; 
\fill (x1t1) circle (2pt);
\fill (x1t2) circle (2pt);
\fill (x2t3) circle (2pt);
\fill (x2t4) circle (2pt);
\draw [directed] (x1t1) -- (x2t3);
\draw [directed] (x1t2) -- (x2t4);
\draw [dashed,thick] (x1t1) -- (x1t2);
\draw [dashed,thick] (x2t3) -- (x2t4);
\draw [enddirected]  (x2t3)--(1.75,0);
\draw [directed]  (x2t4)--(x1t1);
\end{tikzpicture}}}}= -  \Delta''(0) \Delta'(0^+) I_{\rm ss}\\
\ca F_2 &=& {\parbox{2cm}{{\begin{tikzpicture}
\coordinate (x1t1) at  (0,0) ; 
\coordinate (x1t2) at  (0,.75) ; 
\coordinate (x2t3) at  (1.25,0) ; 
\coordinate (x2t4) at  (1.25,0.75) ; 
\fill (x1t1) circle (2pt);
\fill (x1t2) circle (2pt);
\fill (x2t3) circle (2pt);
\fill (x2t4) circle (2pt);
\draw [directed] (x1t1) -- (x2t3);
\draw [directed] (x1t2) -- (x2t4);
\draw [dashed,thick] (x1t1) -- (x1t2);
\draw [dashed,thick] (x2t3) -- (x2t4);
\draw [enddirected]  (x2t3)--(1.75,0);
\draw [directed] (x2t4) arc(90:-90:0.375);
\end{tikzpicture}}}} =   -\Delta (0) \Delta'''(0^+) I_1 I_{\rm tp}\\
\ca F_3 &=& {\parbox{2cm}{{\begin{tikzpicture}
\coordinate (x1t1) at  (0,0) ; 
\coordinate (x1t2) at  (0,.75) ; 
\coordinate (x2t3) at  (1.25,0) ; 
\coordinate (x2t4) at  (1.25,0.75) ; 
\fill (x1t1) circle (2pt);
\fill (x1t2) circle (2pt);
\fill (x2t3) circle (2pt);
\fill (x2t4) circle (2pt);
\draw [directed]  (x2t3)--(x1t1) ;
\draw [directed] (x1t2) -- (x2t4);
\draw [dashed,thick] (x1t1) -- (x1t2);
\draw [dashed,thick] (x2t3) -- (x2t4);
\draw [enddirected]  (x1t1)--(-.5,0);
\draw [directed]  (x2t4)--(x1t1);
\end{tikzpicture}}}} = 0 \mbox{~~~~(see below)} \\
\ca F_4 &=& {\parbox{2.5cm}{{\begin{tikzpicture}
\coordinate (x1t1) at  (0,0) ; 
\coordinate (x1t2) at  (0,.75) ; 
\coordinate (x2t3) at  (1.25,0) ; 
\coordinate (x2t4) at  (1.25,0.75) ; 
\fill (x1t1) circle (2pt);
\fill (x1t2) circle (2pt);
\fill (x2t3) circle (2pt);
\fill (x2t4) circle (2pt);
\draw [directed]  (x2t3)--(x1t1) ;
\draw [directed] (x1t2) -- (x2t4);
\draw [dashed,thick] (x1t1) -- (x1t2);
\draw [dashed,thick] (x2t3) -- (x2t4);
\draw [enddirected]  (x1t1)--(-.5,0);
\draw [directed] (x2t4) arc(90:-90:0.375);
\end{tikzpicture}}}}= -\Delta'(0^+)\Delta''(0) I_1 I_{\rm tp}\qquad  \\
\ca F_5 &=& \half~ {\parbox{2cm}{{\begin{tikzpicture}
\coordinate (x1t1) at  (0,0) ; 
\coordinate (x1t2) at  (0,.75) ; 
\coordinate (x2t3) at  (1.25,0) ; 
\coordinate (x2t4) at  (1.25,0.75) ; 
\fill (x1t1) circle (2pt);
\fill (x1t2) circle (2pt);
\fill (x2t3) circle (2pt);
\fill (x2t4) circle (2pt);
\draw [directed] (x1t1) -- (x2t4);
\draw [directed] (x1t2) -- (x2t4);
\draw [dashed,thick] (x1t1) -- (x1t2);
\draw [dashed,thick] (x2t3) -- (x2t4);
\draw [directed] (x2t3) arc(-90:90:0.375);
\draw [enddirected]  (x2t4)--(1.75,0.75);
\end{tikzpicture}}}} = \half\Delta (0) \Delta'''(0^+) I_1 I_{\rm tp}\\ 
\ca F_6 &=& \half~{\parbox{2.1cm}{{\begin{tikzpicture}
\coordinate (x1t1) at  (0,0) ; 
\coordinate (x1t2) at  (0,.75) ; 
\coordinate (x2t3) at  (1.25,0) ; 
\coordinate (x2t4) at  (1.25,0.75) ; 
\fill (x1t1) circle (2pt);
\fill (x1t2) circle (2pt);
\fill (x2t3) circle (2pt);
\fill (x2t4) circle (2pt);
\draw [directed] (x1t1) -- (x2t4);
\draw [directed] (x1t2) -- (x2t4);
\draw [dashed,thick] (x1t1) -- (x1t2);
\draw [dashed,thick] (x2t3) -- (x2t4);
\draw [enddirected]  (x2t3)--(1.75,0);
\draw [directed] (x2t4) arc(90:-90:0.375);
\end{tikzpicture}}}} = \half \Delta (0) \Delta'''(0^+) I_1 I_{\rm tp} \\
\ca F_7 &=& {\parbox{2.5cm}{{\begin{tikzpicture}
\coordinate (x1t1) at  (0,0) ; 
\coordinate (x1t2) at  (0,.75) ; 
\coordinate (x2t3) at  (1.25,0) ; 
\coordinate (x2t4) at  (1.25,0.75) ; 
\fill (x1t1) circle (2pt);
\fill (x1t2) circle (2pt);
\fill (x2t3) circle (2pt);
\fill (x2t4) circle (2pt);
\draw [directed]  (x2t4)--(x1t1) ;
\draw [directed] (x1t2) -- (x2t4);
\draw [dashed,thick] (x1t1) -- (x1t2);
\draw [dashed,thick] (x2t3) -- (x2t4);
\draw [enddirected]  (x1t1)--(-.5,0);
\draw [directed] (x2t3) arc(-90:90:0.375);
\end{tikzpicture}}}} = \Delta'' (0) \Delta'(0^+) I_1 I_{\rm tp}
\eea
\begin{figure}[t]
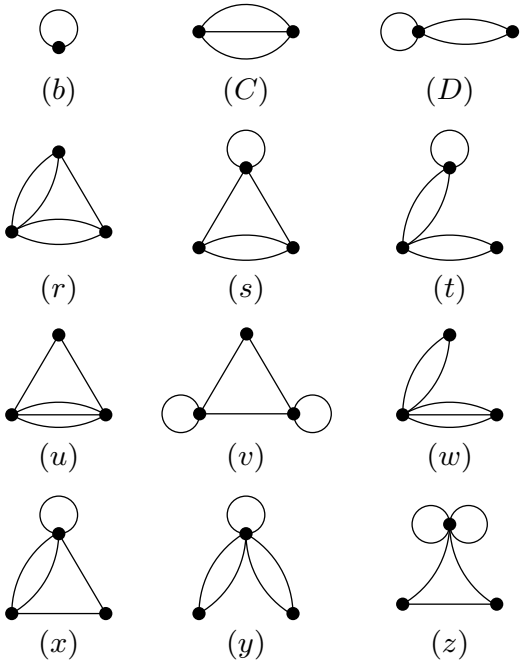

\scalebox{1.25}{$
\begin{array}{ccccccc}
\ITP & &\Isunset & &\ItwoloopTP \\
(b) && (C) && (D) \medskip\\ 
\Ir &\quad&   \Is   &\quad& \It  \\
(r)  & & (s)  & & (t)  
\medskip\\ 
  \Iu   &\quad& \Iv   &\quad&   \Iw  \\
(u)  & & (v)  & & (w)    \medskip \\
 \Ix    &\quad&  \Iy   &\quad& \Iz  \\
(x)  &\quad& (y)  &\quad& (z)  
\end{array}
$}
\caption{All spatial diagrams for corrections of $F_{\rm c}$ and $\eta$; the first three diagrams (without label) are the 1- and 2-loop contributions. The remaining nine diagrams $(r)$ to $(z)$ are 3-loop contributions.}
\label{f:fc}
\end{figure}%
The non-trivial diagram $\ca F_3$ is 
\bea
&& \ca F_3 =\Delta' (0^+) \Delta''(0^+) \int_{k,q}\int_{t_1,t_2,t_3} \nn\\
&& \qquad  \quad \times \rme^{-k^2 t_1- (k+q)^2 t_2 - q^2 t_3 - m^2(t_1+t_2+t_3)} \mbox{sign} (t_3-t_1) \nn\\
&&= \int_{k,q} \frac{k^2-q^2}{\left(k^2{+}m^2\right)
    (q^2{+}m^2 )  (k^2{+}q^2  {+} 2
   m^2)
    ((k{+}q)^2{+}m^2 )} \nn\\
&&=0    .
\eea
Further cancelations read
\be
\ca F_2+\ca F_5+\ca F_6 = 0, \quad \ca F_4 + \ca F_7 = 0.
\ee
Thus 
\be
\sum_{i=1}^7 \ca F_i  = \ca F_1 = -\Delta'(0^+) \Delta''(0) I_{\rm ss}.
\ee
The {\em sunset diagram} $I_{\rm ss}$ reads
\bea
I_{\rm ss} &=& \Isunset\nn\\
 &=& \int_{k,p} \frac1{(k^2+m^2)(p^2+m^2)[(k+p)^2+m^2]}.\qquad 
\eea
It is  evaluated in appendix \ref{s:The sunset diagram}, 
\be
I_{\rm ss} = m^2\left[-\frac{3}{2
   \epsilon ^2}-\frac{9}{4 \epsilon }-\frac{3(7-4 C_3)}{8}  + ... \right] (\epsilon I_1)^2. 
\ee

\subsection{3 loop}
At 3-loop order there are nine diagrams, shown on Fig.~\ref{f:fc}. 
Diagram $(r)$ reads
\be
 {(r)}=  \Delta '(0^+) \left[3 \Delta ''(0)^2+2 \Delta
   '''(0^+) \Delta '(0^+)\right]\Ir. 
\ee
The integral is calculated in appendix \ref{a:Ir}, 
\bea
I_r&=& \Ir \nn\\
&=& m^{2} (\epsilon I_1)^3\left[-\frac{1}{\epsilon ^3}-\frac{17}{6 \epsilon
   ^2}+\frac{36 C_3-67}{12 \epsilon }+\ca O(\epsilon^0) \right]. \quad 
\eea
After some tedious calculations, one surprisingly finds that  
all remaining contributions vanish, 
\be
{(s)}={(t)} = {(u)} = {(v)} = {(w)} = {(x)}
= {(y)} = {(z)} = 0.
\ee

\subsection{Critical force to 3-loop order and flow-equation}\label{secD}
Up to UV-cutoff dependent terms, 
\bea
\label{Fc}
f_{\rm c} &=& \Delta_0'(0^+) I_{\rm tp} - \Delta_0'(0^+)  \Delta_0''(0^+) I_{\rm ss} \nn\\
&&+  \left[2 \Delta_0 '''(0) \Delta_0 '(0)^2+3
   \Delta_0 '(0) \Delta_0 ''(0)^2\right]I_r + ...\qquad 
\eea
The following flow is finite (the $\Lambda$-dependent terms have disappeared under the $m$ variation)
\bea\label{99}
&& - m{\partial_m}  \left< u-w \right> \equiv  m{\partial_m}  \left[ f_{\rm c} m^{-2} \right] \nn\\
&& =- \epsilon m^{-2}\Big\{ \Delta_0'(0^+) I_{\rm tp} -2   \Delta_0'(0^+)  \Delta_0''(0^+) I_{\rm ss} \nn\\
&&\qquad~~~~~~~+ 3   \big[2 \Delta_0 '''(0) \Delta_0 '(0)^2+3
   \Delta_0 '(0) \Delta_0 ''(0)^2\big]I_r  \nn\\ 
   && \qquad~~~~~~~+ \ca O(\Delta_0^4) \Big\}
\eea
The following step is to  replace $\Delta_0(w)$ by $\Delta_{\rm eff}(w)$ using \Eq{Delta0-to-Delta-eff}.
In the next step, we   use  a generalization of \Eq{21}  
\be\label{100}
 \Delta_{\rm eff}(w) = \frac{1}{\epsilon I_1 }\Big(\frac{\rho_m}{\tilde \rho}\Big)^2 \tilde \Delta(  w \tilde \rho/\rho_m ),
\ee
where 
\be\label{rhom}
  \rho_m := \lim_{w\to 0}\frac{\Delta_{\rm eff}(w)}{|\Delta_{\rm eff}'(w)|}, \quad   \tilde \rho := \frac{\tilde \Delta(0)}{|{\tilde \Delta}'(0)^+|}.
\ee 
The ratio $\rho_m$ can be measured in simulations and experiments. 
The   combination $\tilde \rho$ is a theoretical object,   depending on the choice 
of scheme to solve the FRG-equation, see the ansatz \eq{Delta-ansatz}.

For the perturbative calculation of $f_{\rm c}$, there are two important points: First, the integrals $I_{\rm tp}$, $ I_{\rm ss}$ and $I_{\rm r}$   can be combined  into
 the dimensionless combinations
\bea
\frac{ I_{\rm tp}}{m^2 \epsilon I_1} &=& \frac{2}{(\epsilon -2) \epsilon } \\
\frac{ I_{\rm ss}}{m^2 (\epsilon I_1)^2} &=&  -\frac{3}{2
   \epsilon ^2}-\frac{9}{4 \epsilon } -\frac{3}{8} \left(7-4 C_3\right)+ ...\\
   \frac{ I_{\rm r}}{m^2 (\epsilon I_1)^3} &=& -\frac{1}{\epsilon
   ^3}-\frac{17}{6 \epsilon ^2}+\frac{36 C_3-67}{12 \epsilon }+...
\eea
Second, a global factor of $\rho_m/\tilde \rho$   appears from the single $\Delta'(0^+)$, whereas $\Delta''(0)$ and $\Delta'(0^+)\Delta'''(0^+)$ do not give additional factors.
Therefore \Eq{99}, expressed in terms of the renormalized dimensionless disorder $\tilde \Delta$, and scales $m$ and $\rho_m$, reads
\bea
&& m{\partial_m}  \left[ f_{\rm c} m^{-2} \right] \nn\\
&&=\frac{2 }{2-\epsilon} \frac{\rho_m}{\tilde \rho}\Big[-\tilde \Delta'(0^+) \nn\\
&& \qquad + 3 \tilde \Delta'(0^+) \tilde \Delta ''(0)\Big(1 + \epsilon 
    (1{-}C_3)  +... \Big) \nn\\
&&
\qquad+  (C_3{-}6 ) \tilde \Delta'(0^+) \left(3 \tilde \Delta
   ''(0)^2+2 \tilde \Delta '''(0) \tilde \Delta'(0^+)\right) \nn\\
&& \qquad  + \ca O(\tilde \Delta^4) 
   \Big] \nn\\
&&=    \tilde{\ca A}  \frac{\rho_m}{\tilde \rho}.
\label{115}
\eea
We   grouped all terms for a given loop-order in the same line, and expanded as far as necessary in $\epsilon$.
Inserting the RF fixed point, we find
\bea
 \tilde {\ca A} &=&
 \frac{2 }{2-\epsilon}  \Big[-\tilde \Delta'(0^+) \nn\\
&& \qquad + 3 \tilde \Delta'(0^+) \tilde \Delta ''(0)\Big(1 + \epsilon 
    (1{-}C_3)  +... \Big) \nn\\
&&
\qquad+  (C_3{-}6 ) \tilde \Delta'(0^+) \left(3 \tilde \Delta
   ''(0)^2+2 \tilde \Delta '''(0) \tilde \Delta
   '(0)\right)  \nn\\
&& \qquad  + \ca O(\tilde \Delta^4) 
   \Big] \nn\\
&=& \frac{\epsilon
   }{3}+0.007784584 \epsilon ^2+0.0170387
   \epsilon ^3+\ca O(\epsilon ^4). 
\label{amplitudetildeA}
\eea\begin{figure}[t!]
\Fig{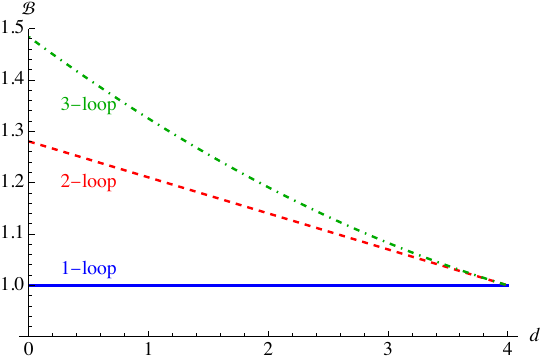}
\caption{The amplitude $ \ca B $ in \Eq{amplitudeB}. 1-loop (blue solid), 2-loop (red, dashed), 3-loop (green, dot-dashed).}
\label{f:amplitudeAovertilderho}
\end{figure}
To solve \Eq{115} we  use that $\rho_m\sim m^{-\zeta}$, to obtain 
\be
\frac{f_{\rm c}} {m^2}  =  -  \frac{\tilde{ \ca A}}{\zeta} \frac{\rho_m}{\tilde \rho} + m\mbox{-independent term}.
\ee
This is equivalent to  
\bea\label{fc}
f_{\rm c} &=& f_0 - \ca B    {\rho_m}  m^{2} + \ca O(m^2), \\
\label{amplitudeB}
\ca B &:=&   \frac{\tilde {\ca A}}{\zeta \tilde \rho } =1+ 0.070061 \epsilon +0.0127138 \epsilon^2+\ca O (\epsilon ^3 ),\qquad  \\
\label{tilderho}
\tilde \rho &=&  1 -0.190020 \epsilon  +0.27397 \epsilon^2 + \ca O(\epsilon^3).
\eea
Note that we added a term $f_0\sim \Lambda^{d-2}\Delta_0'(0^+)$ due to the leading UV divergence of the tadpole diagram \eq{hardcutoff}
which diverges with the UV cutoff $\Lambda$ as   $\Lambda^{d-2}$,  times the bare $\Delta_0'(0^+)$: since this is a strong UV divergence, we 
  used $\Delta'(0^+)$ at the start of the RG flow, i.e.\ the microscopic $\Delta_0'(0^+)$.

\begin{figure}[t]
\Fig{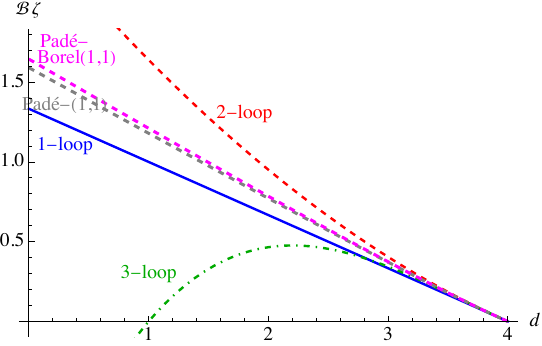}
\caption{The combination $\ca B \zeta \equiv \tilde {\ca A}/\tilde \rho$. 1-loop (blue solid), 2-loop (red, dashed), 3-loop (green, dot-dashed).}
\label{f:amplitudeBzeta}
\end{figure}

We tried resummations for $\tilde {\ca A}$, $\ca B$ and   $\zeta \ca B \equiv \tilde {\ca A}/\tilde \rho$. The series for $\ca B$ has only positive terms, thus the result increases at each order and we do not know how to resum. The combination $\zeta \ca B $ reported on Fig.~\ref{f:amplitudeBzeta} is alternating, and both the diagonal Pad\'e resummation, as the diagonal Pad\'e-Borel resummation lie close to each other and the 1-loop result. We report all 3-loop values in table \ref{tab:cal-B}. The prediction for $\ca B$ from the extrapolation of $\zeta \ca B$ uses the 
best numerically available values for $\zeta$. 
\begin{table}
\begin{tabular}{ |c || c | c|  c| c |}
\hline
   & $d=1$ & $d=2$ & $d=3$  &  $d=4$\\
\hline  \hline 
  $\ca B$ (direct)& $1.32$ & $1.19$ & $1.08$ &  $1$ \\
\hline
$\zeta \ca B$ (Pad\'e-Borel) & $1.21$ & $0.78$ & $0.374$ &  $0$ \\
\hline 
$\ca B$ (using $\zeta \ca B$)& 0.96 & $1.036$ & $1.048$ &  $1$ \\
\hline 
$\ca B$ (estimate and error bars) & 1.3(4) & $1.1(1)$ & $1.06(2)$ &  $1$\\
\hline
$\ca B$ (numerics) & 1.8(2) & - & -  &  - \\
\hline
\end{tabular}
\caption{Values for $\ca B$ using either a direct resummation of $\ca B$ in \Eq{amplitudeB} (first line), or the combination $\zeta \ca B$ (second), which is then divided by the numerically known value of $\zeta$ (third line). 
The fourth line is an estimate, based on the trend of the direct extrapolation, and our lack of confidence in the precision of the results. 
For the numerical value see section \ref{s:Critical force}.}
\label{tab:cal-B}
\end{table}%
We tried to improve the extrapolation by linking to the exactly known critical force in 
dimension $d=0$. As can be seen in appendix \ref{Critical force in d=0}, there is an additional $\ln m$ divergence which prevents from exploiting this result.

\section{Critical force for CDWs}
\label{Critical force for CDWs}
\subsection{Summary of known results}
In \cite{WieseFedorenko2018,WieseFedorenko2019}
it was shown that charge-density waves at depinning map onto the $O(n)$ model in the limit of $n\to -2$. The latter further maps onto loop-erased random walks 
\cite{WieseFedorenko2018,WieseFedorenko2019,HelmuthShapira2020,ShapiraWiese2020}. 
In particular, the dynamic exponent $z$ in CDWs equals the fractal dimension of loop-erased random walks. In $\phi^4$-theory, this fractal dimension is given by the dimension of
the traceless rank-2 tensor, 
\be
\ca T^{ij} = \phi^i \phi^{j}- \delta^{ij} \frac1n  \sum_{k=1}^n (\phi^k)^2.
\ee
An interesting question is whether the critical force also has a representation in $\phi^4$ theory. We   show below that this is the case, and the critical force formally behaves as a logarithmic opeartor in a log-CFT. 

\subsection{Critical force for CDWs}
We first consider the CDW side.
Following the conventions of Ref.~\cite{WieseFedorenko2018}, we parameterize the 
disorder-force correlator $\Delta(u)$ for  CDWs as 
\be\label{104}
\Delta(u) = \Delta(0) -\frac g 2 u (1-u).
\ee
The fixed point for the $\beta$-function \eq{beta-3-loop} is 
\bea
\Delta(0) &=& \frac{\epsilon }{36}+\frac{\epsilon
   ^2}{108}-\frac{\epsilon ^3}{648} \left(1+18 C_3\right)
   + \ca O (\epsilon ^4 ) ,\qquad \\
g&=& \frac{\epsilon }{3}+\frac{2 \epsilon
   ^2}{9}+\frac{ \epsilon ^3}{9} \big[1-2 C_3-2
   \zeta (3)\big]+\ca O ( \epsilon^4).\qquad
   \label{106}
\eea
Thus $\Delta'''(0^+)=0$, and \Eq{Fc} for the critical force simplifies to  
\be
f_{\rm c}   = \frac g 2 \ITP   - \frac {g^2}2 \Isunset +\frac32 g^3 \Ir + ...\label{Fc-diagrammatric}
\ee

\subsection{$\Gamma^{(2)}$ as a function of $n$}
The vector $\phi^4$-theory related to CDWs is  \cite{WieseFedorenko2018}
\be\label{108}
\ca S[\vec \phi] = \int_x \frac12 \left[  \nabla \vec \phi(x)^2\right] + \frac{m^2}2 \vec\phi(x)^2 + \frac g 8 \left[ \vec \phi(x)^2\right]^2 .
\ee
In these conventions, comparable quantities are related, e.g.\ the coupling constants $g$
in \Eqs{104} and \eq{108} are identical. Using the same RG scheme, also all RG functions, and the coupling at the fixed point given in \Eq{106} are identical. 
In this framework, we now evaluate the effective action,  
$\Gamma[\phi] =\frac{m^2}2  \phi^2 + \ca O(g)$, equivalent to  $\Gamma^{(2)}  = m^2 + \ca O(g)$. The result up to 4-loop order reads  
\bea
&& {\Gamma^{(2)}} - m^2= -\frac{g(n+2)}2 \ITP \\\
&& + \frac{g^2(n+2)^2}4 \ItwoloopTP + \frac{g^2(n+2)}2 \Isunset \nn\\
&& -\frac{1}{8} g^3 (n+2)^3 \It -\frac{1}{4} g^3 (n+2)^2 \Iu \nn\\
&& -\frac{1}{8} g^3 (n+2)^3 \Iv -\frac{1}{4} g^3 (n+2) (n+8) \Ir\nn\\
&& -\frac{3}{4} g^3 (n+2)^2\Is   \nn\\
&& + \frac{1}{4} g^4 (n+2) (5n+22) \IfourA \nn\\
&& + \frac{1}{8} g^4 (n+2) (n^2 + 6n +20) \IfourB \nn\\
&& + \frac{1}{4} g^4 (n+2) (5n+22) \IfourC \nn\\
&& + \ca O (g^4, (n+2)^2 )
\eea
By inspection one sees that $f_{\rm c}$ is related to the dominant contribution in the limit of 
$n\to -2$, 
\bea
&& -\frac{\partial}{\partial n} {\Gamma^{(2)}}  \Big|_{n=-2} = \frac{g}2 \ITP
- \frac{g^2}2 \Isunset  +\frac{3g^3}{2} \Ir \nn\\
&& -  3{g^4} \IfourA - \frac{3g^4}2 \IfourB -  3{g^4} \IfourC  \nn\\
&& + \ca O(g^5),
\eea
where the 4-loop contribution was not checked at depinning. 
We conjecture that to all orders in perturbation theory
\be\label{138}
{f_{\rm c} = - \frac{\partial}{\partial n} \Gamma^{(2)}   \bigg|_{n=-2}}. 
\ee
We now use that 
\be
\Gamma^{(2)}(m) =\Gamma^{(2)}(1) m^{\frac1\nu+\eta}.
\ee
Since we  retain the coefficient in front of $\phi^2$, the anomalous dimension $\eta$ of the field is taken out. 
\Eq{138} implies that 
\bea
&& m{\partial_m}  \left[ f_{\rm c} m^{-2} \right] =  -\frac{\partial}{\partial n} \left[\frac{m \partial}{\partial m}  \Gamma^{(2)}(1) m^{\frac1\nu+\eta-2}\right]_{n=-2} \nn\\
&& = -\frac{\partial}{\partial n}\left[\left( {\frac1\nu+\eta-2} \right) \Gamma^{(2)}(1) m^{\frac1\nu+\eta-2}\right]_{n=-2} \nn\\
&& = -\frac{\partial}{\partial n}\left[ {\frac1\nu+\eta} \right]_{n=-2} .
\label{125}
\eea
Let us see where these contributions come from in the RG.
According to Ref.~\cite{KompanietsWiese2019}, 
\bea
\!\!\!\frac1\nu +\eta &=& 2 + \gamma_1 \qquad ~~~~~\mbox{(eq.~(19) of \cite{KompanietsWiese2019})},\\
\gamma_1 &=& \mu \partial_\mu \ln Z_1 \qquad \mbox{(eq.~(14) of \cite{KompanietsWiese2019})},\\
\ca S &=& \int_x Z_1 \frac{m^2}{2}\vec \phi(x)^2 + \frac{Z_2}2 [ \nabla \vec\phi(x)]^2 \nn\\
&&+ Z_4 \frac{16 \pi^2}{4!} g \mu^{\epsilon } [\vec \phi(x)^2]^2 \quad \mbox{(eq.~(8) of \cite{KompanietsWiese2019})}.~~~~~~~~~~
\eea
Thus $f_{\rm c}$ is entirely given by the renormalization group factor $Z_1$, and does not invoke a  renormalization of the field. 

Let us finally use \Eq{125}, and the 6-loop results of \cite{KompanietsPanzer2017}. 
\bea
&& m{\partial_m}  \left[ f_{\rm c} m^{-2} \right] =\frac{\epsilon }{6}+\frac{\epsilon
   ^2}{36}+\frac{1}{72} \big[1-8 \zeta (3)\big] \epsilon
   ^3 \label{126}
\nn\\
&&+  \frac{ -70 \zeta (3)+2800 \zeta (5)-6 \pi
   ^4+25  }{6480}\epsilon ^4 \nn\\
&& + \bigg[
\frac{7 \zeta (3)^2}{162}-\frac{115 \zeta
   (3)}{3888}+\frac{29 \zeta (5)}{648}-\frac{833 \zeta
   (7)}{432}+\frac{5 \pi ^6}{8748} \nn\\
&& \qquad -\frac{7 \pi
   ^4}{77760}+\frac{7}{7776}
\bigg]   \epsilon^5 \nn\\
&& + \bigg[
\frac{344 \zeta (3)^3}{729}+\frac{443 \zeta
   (3)^2}{1296}+\frac{953}{486} \zeta (5) \zeta
   (3)+\frac{7 \pi ^4 \zeta (3)}{9720}\nn\\
&&\qquad    -\frac{305 \zeta
   (3)}{23328}+\frac{1511 \zeta
   (5)}{23328}-\frac{17815 \zeta
   (7)}{46656}+\frac{60451 \zeta (9)}{6561} \nn\\
&& +\frac{697
   \text{Z35}}{540}-\frac{168317 \pi
   ^8}{244944000}+\frac{47 \pi ^6}{489888}-\frac{23
   \pi ^4}{93312}-\frac{5}{15552}
\bigg]\epsilon^6  \nn\\
&&+ \ca O(\epsilon ^7).
\eea
We   find that   this agrees up to 3-loop order with the result obtained for depinning of CDWs. 

\begin{figure}
\Fig{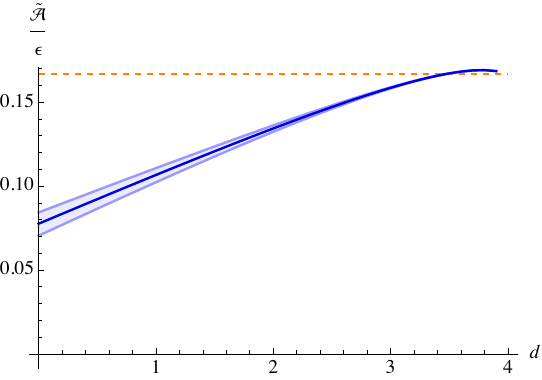}
\caption{$\tilde {\ca A}_{\rm c}/\epsilon$ for CDW, given by \Eq{126}. The error bars are probably an underestimation, as they do not catch the singularity at $d=2$.}
\label{fcOverEpsCDW}
\end{figure}

We finally need to resum this asymptotic series. 
A relevant dimension is $d=3$, for which we find (with possibly strongly underestimated error bars)
\be
-\partial_n \left[\frac1\nu + \eta \right]_{n=-2,\epsilon=1} = 0.1585(5)
\ee
In dimension $d=2$, we can try to use CFT data in \Eq{125}. As we show in appendix 
\ref{CDW-d=2} this expression diverges when taking the limit of $n\to -2$. We conjecture that for $d\le 2$ the critical force acquires an 
additional singularity not captured by the $4-\epsilon$ expansion. 
While our extrapolations are shown on Fig.~\ref{fcOverEpsCDW}   down to $d=0$,   we should thus not trust it for $d\le 2$. 

We saw above that the critical force for CDWs can be calculated in the $O(n)$-model, by deriving $\Gamma^{(2)}$ w.r.t.\ $n$. This means that the operator in question is not living inside the theory at $n=-2$, but in the larger set of theories around $n=-2$. 
This sometimes happens in log-CFTs. Here we give one  prescriptions to obtain $f_{\rm c}$ directly inside the theory 
at $n=-2$,
\be
{-\frac g{6}    \phi_1(x)^3 \rme^{-{\ca S}} \to f_{\rm c} \phi_1(x) }.
\ee
This means to evaluate the insertion $-\frac g{6}    \phi_1(x)^3$ inside the interacting field theory, and retain the perturbative corrections proportional to $\phi_1$; their amplitude is $f_{\rm c}$. This  can be achieved by calculating the 2-point function of $\phi_1(x)^3$ with $\phi_1(y)$. 
The logic behind this and alternative constructions are discussed in appendix \ref{Observables inside the theory $n=-2$}.

\subsection{CDWs and  log-CFT}
We start this section with a reminder of logarithms in self-avoiding polymers \cite{Cardy2013}. The reader not familiar with the subject is invited to consult 
appendix \ref{s:Logarithmic operators for self-avoiding polymers: Considerations from Field Theory} or the original publication \cite{Cardy2013}, where the math is worked out. 
The general idea is that there are two operators $\ca E$ and $ \tilde {\ca E}$, which at a critical value $n_{\rm c}$ of a control parameter $n$ have the same full scaling dimension $x_{\ca E}(n_{\rm c}) = x_{\tilde {\ca E}}(n_{\rm c})$, and moreover become identical as operators. Approaching $n_{\rm c}$, 
there are then two differences (or derivatives) one may consider, the difference between the operators $\ca E$ and $\tilde {\ca E}$, and the difference between their scaling dimensions $x_{\ca E}(n) - x_{\tilde {\ca E}}(n)$. It is a matter of conventions whether these differences vanish or are finite. If they vanish, we should divide by $n-n_{\rm c}$, equivalent to taking a derivative. Let us write the relations in the conventions of appendix \ref{s:Logarithmic operators for self-avoiding polymers: Considerations from Field Theory}, where the differences are finite. Define 
\bea
{\cal C} &:=&\lim_{{n\to n_{\rm c}}} [x_{\ca E}(n)-x_{\tilde{\ca E}}(n)]\ca E  \nn\\
&\equiv& \lim_{{n\to n_{\rm c}}} [x_{\ca E}(n)-x_{\tilde{\ca E}}(n)]\tilde {\ca E}, \qquad  \\
{\cal D} &:=& \lim_{{n\to n_{\rm c}}} \ca E - \tilde {\ca E}.
\eea
In appendix \ref {s:Logarithmic operators for self-avoiding polymers: Considerations from Field Theory} we show that this implies 
\bea
\!\!\!\left< \ca D(0) \ca D(r) \right> & =& -\frac{-2 \alpha \ln (r) + \mbox{const}}{r^{{2 x(0)}}},  \\
\left< \ca C(0) \ca D(r) \right>
&   =& \frac{ \alpha}{r^{{2 x(0)}}}  , \\
\left< \ca C(0) \ca C(r) \right> &=& 0 , \\
\alpha &:=&  A(0)     \Big(x_{{\cal E}}'(0)- x_{\tilde {\cal E}}'(0)\Big)\quad \nn\\ 
&\equiv&  \tilde A(0)     \Big(x_{{\cal E}}'(0)- x_{\tilde {\cal E}}'(0)\Big)\ .
\eea
These relations show that
logarithms in CFTs are rather common, and appear when one considers derivatives of operators w.r.t.\ a control parameter, here $n$. 
This is indeed what has been done in \Eq{138}.

\begin{figure}[t]
\includegraphics[width=\columnwidth]{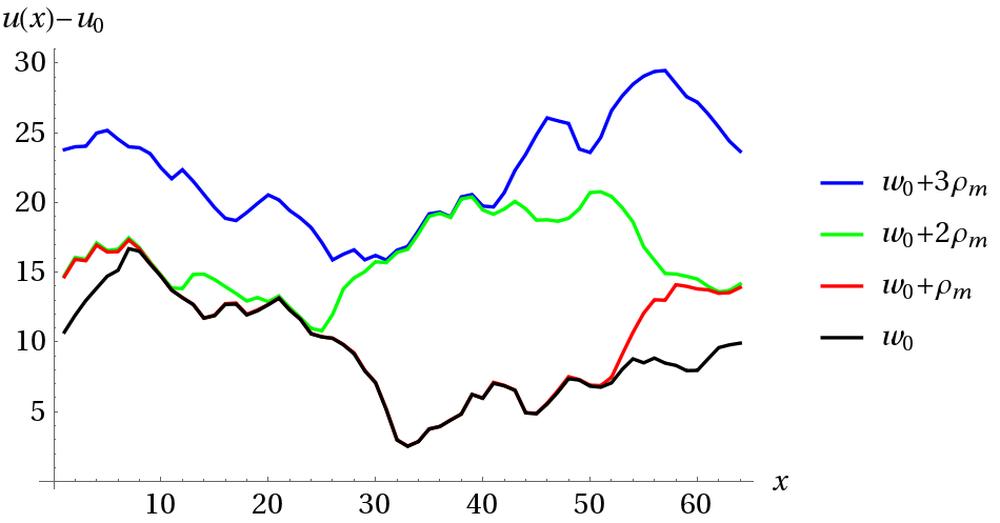}
\caption{$u(x)-u_0$ for $L = 64$, and  $mL=8$. Between successive samples, the control parameter $w$ is increased by   $\rho_m$, starting at $w=w_0$. 
We see that   augmenting $w$ by the correlation length  $\rho_m$ (in $u$-direction), $u(x)$ takes a different configuration at a substantial fraction of sites.} 
\label{fig:z}
\end{figure}

\section{Numerical simulations}
\label{s:simulations}
Let us finally verify our analytical predictions with numerical simulations. 
\subsection{Implementation}
We simulate a discretized version of the equation of motion \eq{EOM} for a string ($d=1$), using  code written in Julia \cite{BezansonEdelmanKarpinskiShah2017}. 
The lattice constant is set to 1, so that   the interface position $u_x\in \mathbb R$ is a vector of   size $L$, with index $x=\{1,...,L\}$.
The random   forces $F(x,u_x)$ are drawn from a Gaussian distribution with mean zero and variance one, independent for each $x$, and $u_x \in \mathbb Z$. For non-integer values of  $u_x$, the force is interpolated linearly between the closest two integer neighbors. The lattice Laplacian is defined by
\begin{equation}\label{Eqns1}
    \nabla^2 u_x := u_{x-1}+u_{x+1}-2u_x, 
\end{equation}
with $u_{0}:=u_L$, and $u_{L+1}:=u_1$. 
The total force acting on site $x$ is   
\begin{equation}\label{Eqns2}
    F_{\rm tot} (u_x) :=m^2\left(w-u_x\right)+\nabla^{2}u_x+F(x,u_x).
\end{equation}
The position $u_x$ of the interface at site $x$ is increased if the 
 force acting on it is   positive. Due to Middleton's theorem \cite{Middleton1992}, to find the pinning configurations, 
 one can move a monomer until the   force acting on it vanishes \cite{RossoKrauth2001b}. This is much more efficient than 
 directly integrating the equation of motion \eq{EOM}. 
 
If   monomer $x$ is at  position  $u \equiv u_x$ we can estimate the  total force acting on it at   position $u +\delta u$ as
\begin{equation}\label{Eqns3}
    F_{\rm tot}({u+\delta u}) = F_{\rm tot}(u)+\frac{\partial  F_{\rm tot}(x,u) }{\partial  u}\delta u  .
\end{equation}
This estimate is valid as long as $u_x+\delta u$ is smaller than the next integer, and we use the right-hand derivative at integer $u_x$. 
In our algorithm $\frac{\partial  F_{\rm tot}(x,u) }{\partial  u} = \partial_{u}{F(x,u)} - m^2 -2 $.
If \Eq{Eqns3} is positive when evaluated at the next integer,  shift $u$ to this value. 
If this is not the case, we   move by $\delta u = - {F_{\rm tot}(u))}/{\frac{\partial F_{\rm tot}(x,u)}{\partial u}}$, s.t.\ at the end of the move $F_{\rm tot}$ vanishes.

\begin{figure}[t]
\includegraphics[width=\columnwidth]{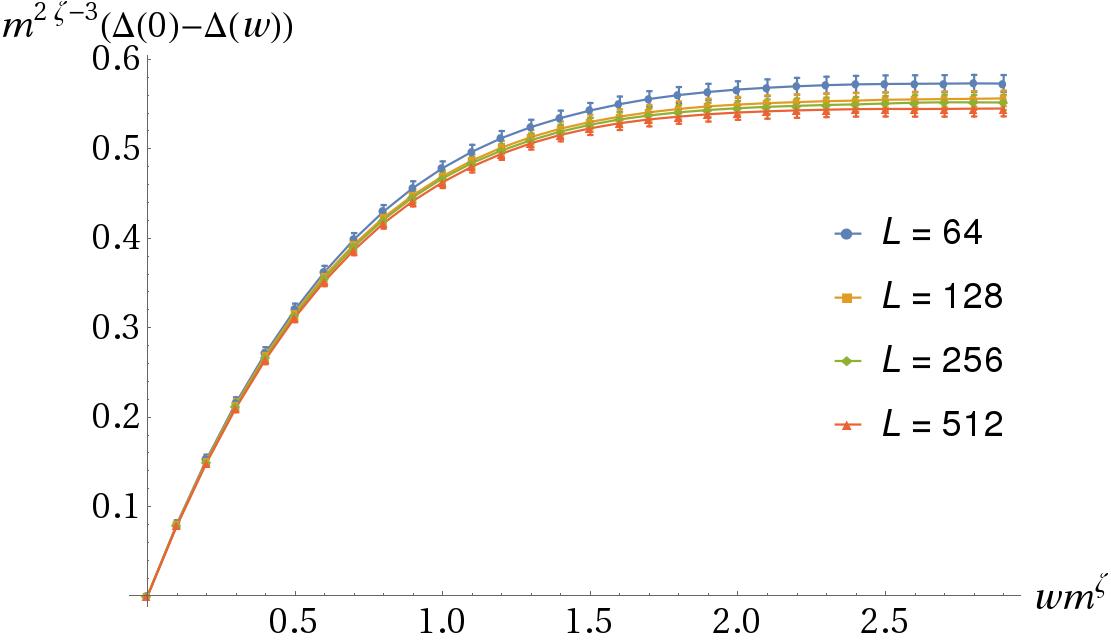}
\caption{$\tilde{\Delta}(0)-\tilde{\Delta}(w)$ for $mL = 4$.}
\label{fig:a}
\end{figure}

\subsection{Measurement of $\Delta(w)$}
\begin{figure}[b]
\includegraphics[width=\columnwidth]{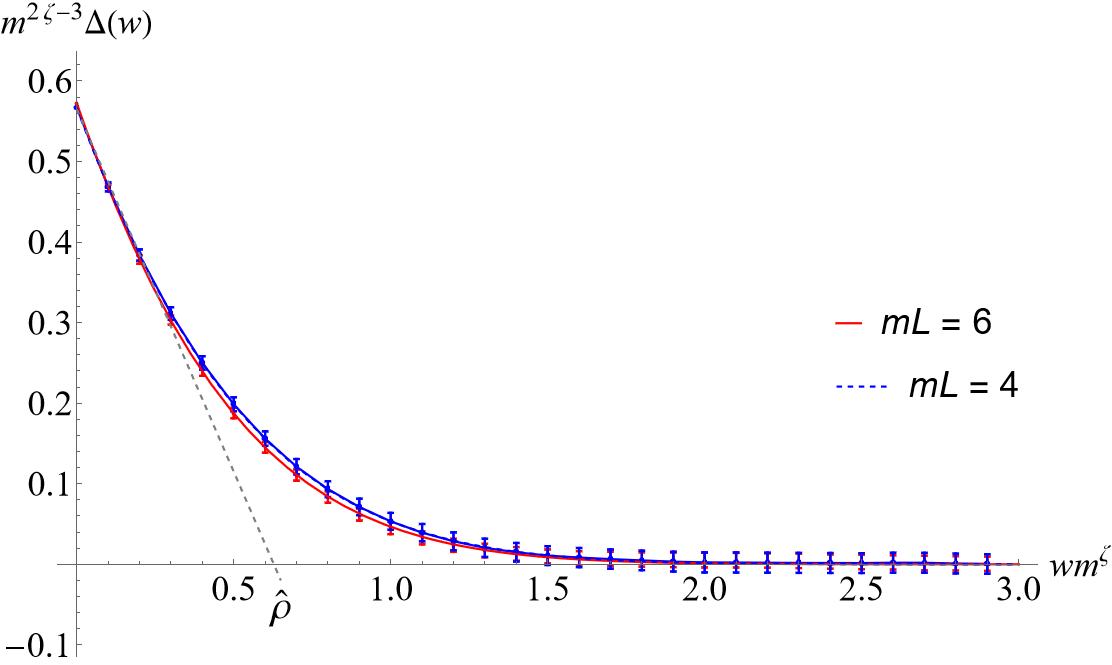}
\caption{$\tilde{\Delta}(w)$ for $mL = 4$ (blue, top curve) and $mL = 6$ (red).}
\label{fig:shape-mL=4-versus-6}
\end{figure}

We measure $\Delta(w)$ and   its second cumulant (variance). In order to get rid of   boundary effects we  need to choose the system size   big enough. From FRG we know that $\Delta(w)$  becomes independent of $L$ in the limit $mL\rightarrow \infty$. In that limit, the spatial correlation function $\overline{[u(x)-u(y)]^2}$
decays exponentially as $  \rme^{-m|x-y|}$ which we associate with
a correlation length $\xi = \frac{1}{m}$. This means that
\bea\label{Eqns4}
    \Delta(w,w') &:=& \frac{m^4}{L^{d}} \int_x \int_y \overline{[u_w(x)-w][u_{w'}(y)- w']}^{\rm c} \nn \\
    &=& m^4  \int_y \overline{[u_w(x)-w][u_{w'}(y)-w']}^{\rm c} \nn\\
    &\approx & m^4 \xi^d \overline{[u_w(x)-w][u_{w'}(x)-w']}^{\rm c} ,
\eea
where we used the spatial exponential decay with correlation length $\xi\ll L$. Since  the disorder forces $F(x,u)$  are statistically invariant under translations in $u$, \Eq{Eqns4} only depends on $|w-w'|$, and we write it as 
\be
\Delta(w-w'):= \Delta (w,w') .
\ee
We  see that \Eq{Eqns4} does not depend on $L$, as long as $\xi \ll L$. As we   saw in the analytic part, and will later confirm in the simulations, 
the function $\Delta(w)$ decays itself approximately exponentially,  $\Delta(w)\approx \Delta(0) \rme^{-\rho_m w}$, which allows us to define an {\em effective disorder correlation length} $\rho_m$ by (see Fig.~\ref{fig:shape-mL=4-versus-6})
\be\label{rho-m}
\rho_m := \frac{\Delta(0)}{|\Delta'(0^+)|}. 
\ee
This is close to the   more natural looking   definition   
\be\label{rho-m'}
\rho_m' := \frac{\int_{w}w \Delta(w)}{\int_{w} \Delta(w)}. 
\ee
We use the definition \eq{rho-m} rather than \eq{rho-m'} for two reasons: First, the latter is difficult to use analytically due to the integral;  second in simulations or experiments the tail of $\Delta(w)$ has large statistical errors,  which gives a large overall error for $\rho_m'$. 

\begin{figure}[t]
\includegraphics[width=\columnwidth]{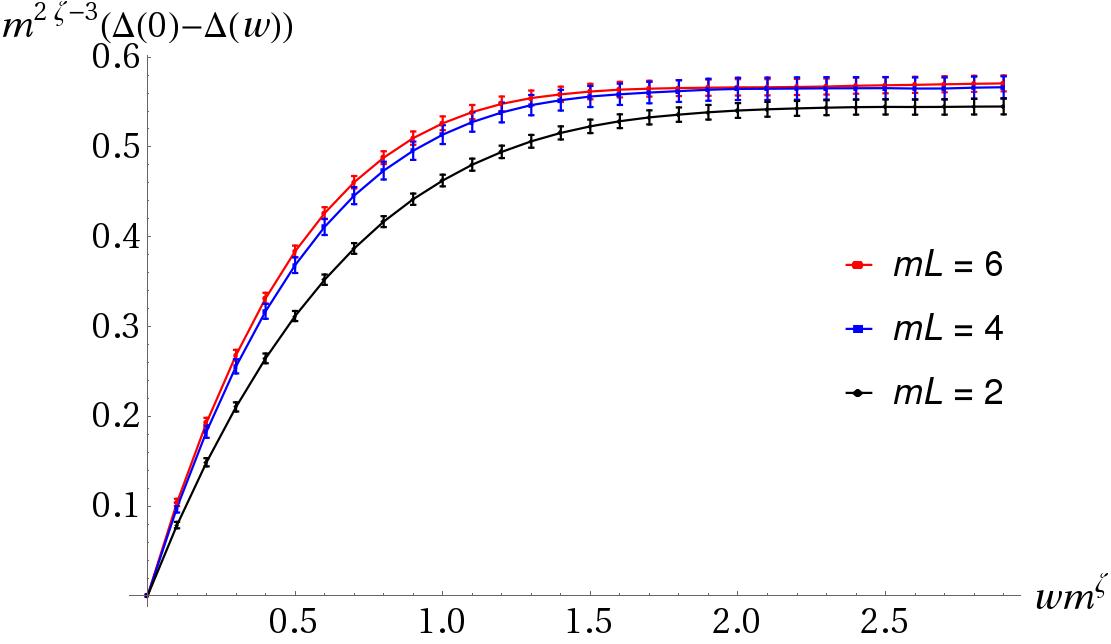}
\caption{$\tilde{\Delta}(0)-\tilde{\Delta}(w)$ for $mL=2$, $4$ and $6$, in a system of size $L=1024$.}
\label{fig:g}
\end{figure}

The variance of $\Delta: =\Delta(u)$, which quantifies the statistical error, can (for each $u$) be estimated from 
\begin{equation}\label{Eqns5a}
 \mbox{var} (\Delta) :=   \frac{\overline{\Delta^2}-\overline{\Delta}^2}{N} ,
\end{equation}
where $N$ is the number of independent samples. 

In order to use \Eq{Eqns5a} we need to get rid of statistically dependent samples. This is achieved by using 
\be
 \mbox{var} (\Delta) := \frac{\overline{\Delta^2}-\overline{\Delta}^2}{N_{\rm eff}}, \quad N_{\rm eff} \approx {\frac{N}{3\rho/\delta w}}, 
\ee 
where $N$ is the number of samples, $\delta w$ the step-size in the simulation between samples taken for $\Delta$, and $\rho_m$ the correlation length defined in \Eq{rho-m}. This is a conservative estimate, assuming that a new independent sample is generated if $w$ is advanced by $3 \rho_m$. It can indeed be seen on the example of Fig.~\ref{fig:z} that when advancing $w$ to $w+3 \rho_m$, the whole line has moved, which reinforces this argument.

\begin{figure}[t]    
\includegraphics[width=\columnwidth]{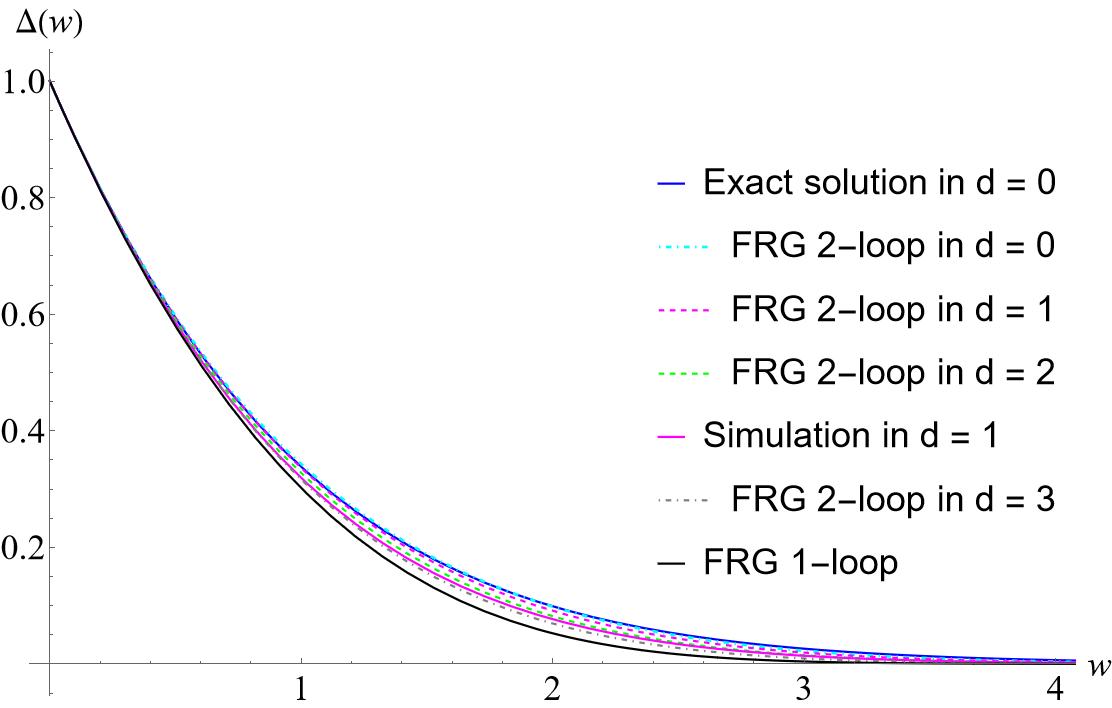}
\caption{Shape comparison: 1-loop FRG   (black, bottom curve), exact solution in dimension $d=0$ (blue solid, top curve), 
Pad\'e resummed 2-loop result from \Eq{Pade-Delta} in $d=0$ for $\alpha=0.35$ (cyan, dotted), 
the same Pad\'e in $d=1$ (magenta, dashed), the same Pad\'e in $d=2$ (green, dashed), our simulations in    dimension $d=1$ (magenta, solid), 
the same Pad\'e in $d=3$ (gray, dot-dashed).}
\label{fig:b}
\end{figure}

In order to show explicitly in the simulations the independence of $ \Delta(w)$ on $L$ in the limit  of large $mL$, we need to eliminate the factors of $m$. By definition,  $u_w(x)-w$ scales as $m^{-\zeta}$,  and $\Delta(w)\sim m^{4-d} u^2 \sim m^{4-d-2\zeta}$, where we used  $\xi = 1/m$ in \Eq{Eqns4}.
This allows us to define the dimensionless correlator $\tilde\Delta(w )$,  
\begin{equation}\label{Eqns6}
  \tilde{\Delta}(w ) := m^{d-4+2\zeta}   \Delta(w m^{\zeta}). 
\end{equation}
Note that this definition is not unique, as one can rescale $\tilde \Delta (w) \to \lambda^{-2} \tilde \Delta(w \lambda)$.

Before considering the results of the numerical simulations, there is a last point we need to address: \Eq{Eqns4} contains a connected average, so one should measure $\overline{u_w(x)-w}$ first. This can be avoided, by sampling the combination 
\be
\Delta(0) - \Delta(w) = \frac12\frac{m^4}{L^{d}} \int_x \int_y \overline{[u_w(x)-w-u_{w'}(y)+ w']^2}^{\rm c}.
\ee
It is this combination we   display   on  Figs.~\ref{fig:a} and~\ref{fig:g}. 
Fig.~\ref{fig:a} shows that the limit of $L\to \infty$ exists, at fixed $mL$. While a system of size $L=64$ is certainly too small,  $L=512$ is large enough to exhibit this limiting behavior. Fig.~\ref{fig:g} analyzes what happens when $mL$ is taken larger, at fixed system size $L=1024$.  The conclusion is that  one should use $mL\ge 6$ to have negligible finite-size effects, physically caused by system-spanning avalanches.

\begin{figure}[t]    
\includegraphics[width=\columnwidth]{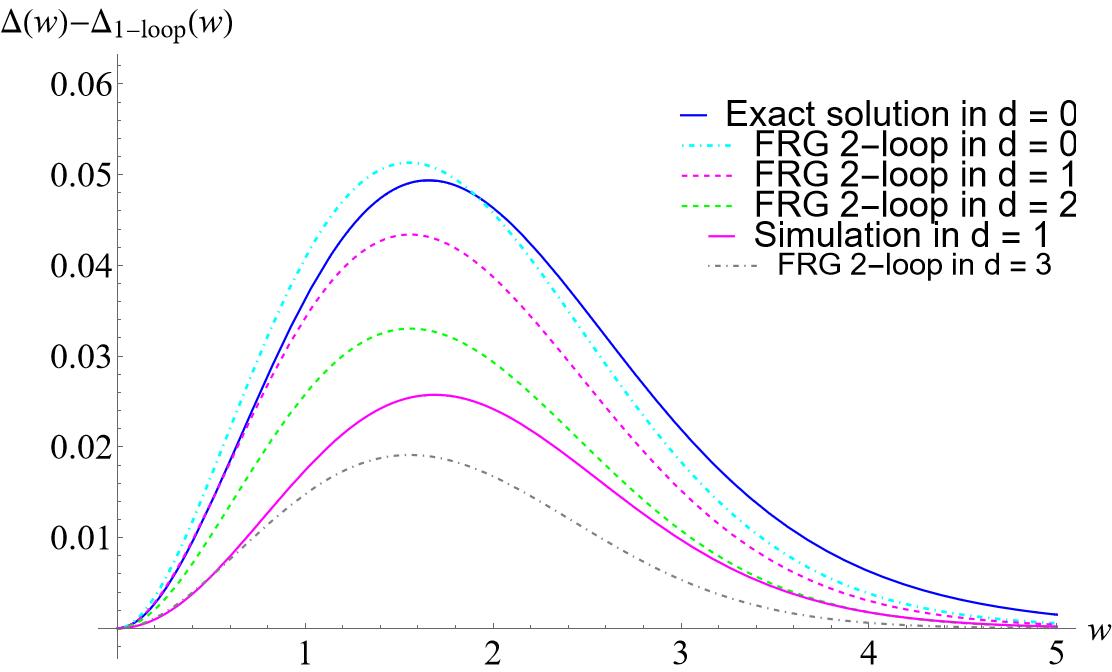}
\caption{Shape comparison of $\Delta(w)-\Delta_{\text{1-loop}}(w)$, with colors as in Fig.~\ref{fig:b}.}
\label{fig:bb}
\end{figure}

\subsection{Comparison of $\Delta(w)$ to the theory}
We now compare the shape of $\Delta(w)$ obtained from simulations to results from  field theory.  
This is delicate as a direct $\epsilon$-expansion is badly converging. At 2-loop order,  we can use a Pad\'e approximant, 
\bea\label{Pade-Delta}
\Delta(w) &=& \epsilon\Delta_1(w) +\epsilon^2  \Delta_2(w) + \ca O(\epsilon^3) \nn\\\
&=& \epsilon  \frac{\Delta_1(w)+\alpha\epsilon\Delta_2(w)}{1+  \epsilon (\alpha-1)\frac{\Delta_2(w)}{\Delta_1(w)}} +  \ca O(\epsilon^3) .
\eea
Our strategy is to use  $\alpha$   to improve convergence; more specifically, we choose $\alpha$, s.t.\ in $d=0$ we recover as precisely as 
possible the exact solution of \cite{LeDoussalWiese2008a}.  As can be seen on Fig.~\ref{fig:b}, this is achieved for $\alpha = 0.35$. Using this value of $\alpha$, we predict the shape of $\Delta(w)$ in $d=1$, see Fig.~\ref{fig:b}. 
This approach works well at two loops for which it was  used in \cite{terBurgBohnDurinSommerWiese2021}. 
In contrast, we were not able to properly resum the $\epsilon$-expansion for $\Delta(w)$ at 3-loop order. Our failed attempts, using Pad\'e resummation and rescaling invariance for optimization, are documented in appendix \ref{f:3-loop resumation}. 

We finally   compare our simulation result (for $mL = 6$, $L = 1024$) in dimension $d = 1$ to the simulation results from Ref.~\cite{MukerjeeWiese2022} in $d = 1$ ($L = 8192$). As can be seen on Fig.~\ref{fig:b} both simulations agree well. We also show simulation results in $d=2$. From experiments \cite{terBurgBohnDurinSommerWiese2021} and continuity of the curves, we expect $\Delta(w)$ in $d=2$ to lie between its counterparts in dimensions $d=1$ and $d=4$. As figure \ref{fig:b} shows, this does not seem to be the case. We expect the system size used in Ref.~\cite{MukerjeeWiese2022} to be too small for $\Delta(w)$ to be in the asymptotic regime.

\begin{figure}[t]
\includegraphics[width=\columnwidth]{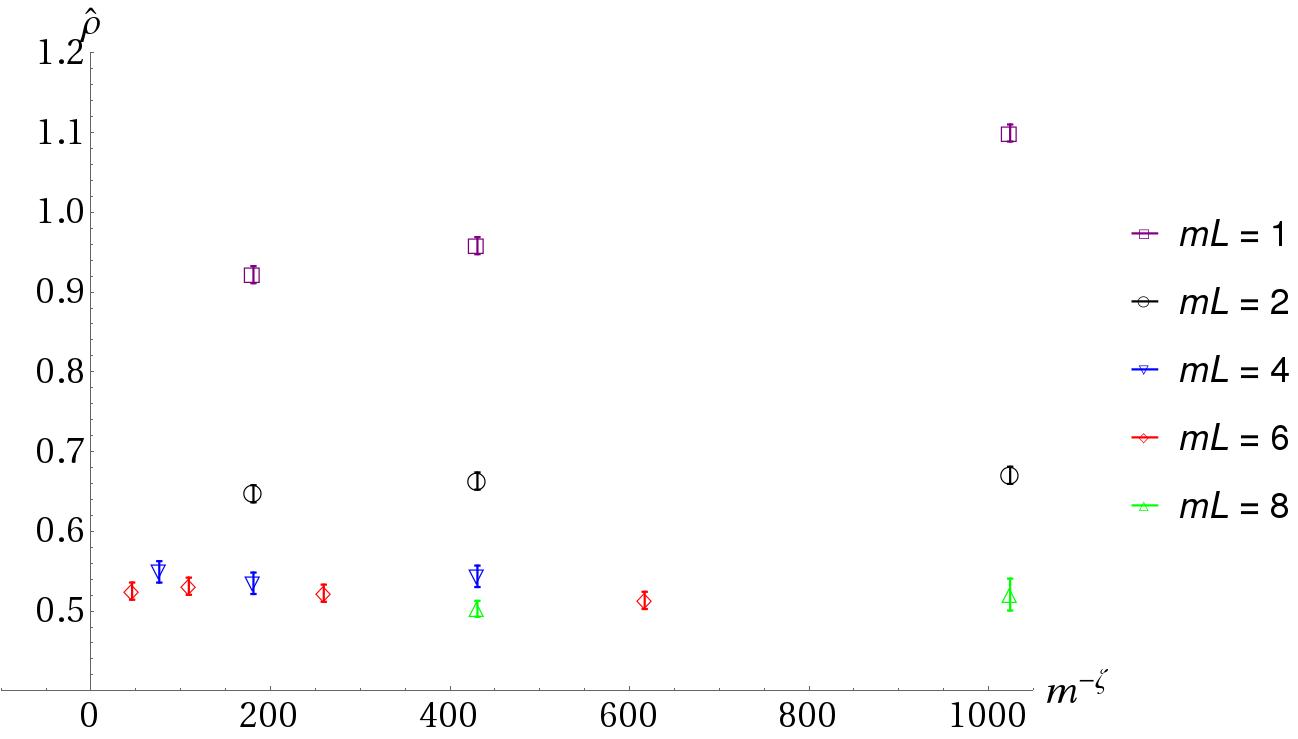}
\caption{$\hat{\rho}$ as a function of  $m^{-\zeta}$, for different $mL$, and different system sizes. The larger systems are to the right.}
\label{fig:rho-hat}
\end{figure}

\subsection{Critical force}
\label{s:Critical force}
\begin{figure}[t]
\includegraphics[width=\columnwidth]{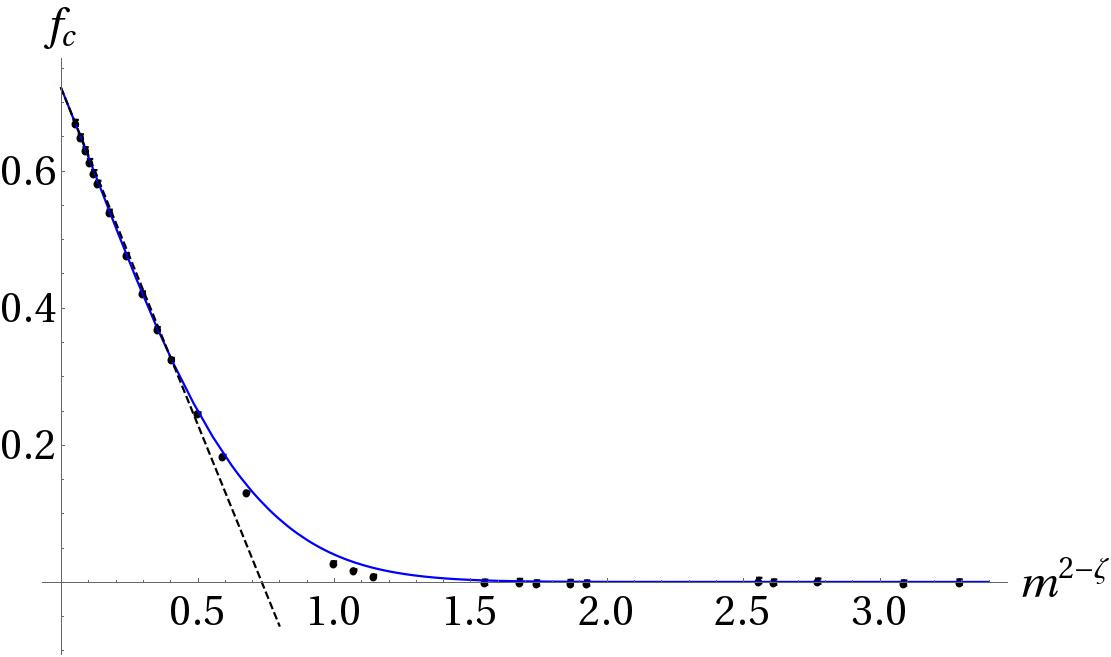}
\caption{$f_{\rm c}$ as a function of $m^{2-\zeta}$ for $mL = 4$. The fit  used to   extrapolate to $m=0$ is via an exponential function, $f_{\rm c} = f_{\rm c}^0 \rme^{-b m^{2-\zeta}} $, with the two fit parameters $f_{\rm c}^0$ and $b$.
The slope indicated with a dashed line is minus $b\equiv \ca B \hat \rho$ as given in \Eq{slope}.}
\label{fig:d}
\end{figure}

We finally compare predictions to simulations for the critical force, defined as 
\begin{equation}\label{Eqns7}
    f_{\rm c} = m^2\overline{(w-u_w)}.
\end{equation}
For large enough $mL$, it does not depend on $L$ for the same reasons as $\Delta(w)$.
\Eq{amplitudeB} predicts   that
\begin{equation}\label{Eqns8}
     {f_{\rm c}}  = f_{\rm c}^0  - \mathcal{B} m^2\rho_m + \ca O(m^2),
\end{equation}
where $\rho_m$ is 
\begin{equation}\label{Eqns9}
    \rho_m = \frac{\Delta(0)}{|\Delta'(0^+)|} =:  \hat{\rho}\, m^{-\zeta}.
\end{equation}
In order to find $\hat{\rho}$ (a numerical value of the simulation) we    plot $\hat \rho = \rho_m m^{\zeta}$, which we    evaluate for small $m$.
 On Fig.~\ref{fig:rho-hat} we find that 
\be\label{hat-rho-sim}
 \hat\rho  =  0.531\pm 0.05  .
\ee
On Fig.~\ref{fig:d} we then plot $f_{\rm c} m^{\zeta-2}$ against $ m^{\zeta-2}$, which yields
\be\label{slope}
b\equiv \ca B \hat \rho = 0.970\pm 0.05. 
\ee
As we   see on Fig.~\ref{fig:d}, for   small     $m$ the critical force $f_{\rm c}(m)$ depends linearly on $m^{2-\zeta}$. From that we   deduce that the $\ca O(m^2)$ in \Eq{fc} is seemingly very small or absent.
Together with the results for $\hat \rho $ shown on Fig.~\ref{fig:d}, this gives our final result for $\ca B$ as 
\be\label{calB-num-final}
\mathcal{B} =  {1.8}\pm0.2 .
\ee
This is in reasonable agreement with the values reported in section \ref{secD}.
Note that $\ca B$ does not depend on the elastic coefficient $c$. This is demonstrated in the appendix \ref{g:B and c}.

\section{Conclusions}
\label{Conclusions}
In this work we   calculated the roughness exponent $\zeta$ to 3-loop order.  Using analytic information in dimension $d=0$ and   Borel resummation allows us to give excellent values for the roughness $\zeta$ in all dimensions, including $d=1$. The predictive power for the shape of the renormalized disorder correlator is weaker: we estimate it to be good    in dimensions $d=3$,   satisfactory  in $d=2$, but insufficient in dimension $d=1$. It is not clear how to implement  a Borel resummation for a whole function.  

We further considered the critical force at depinning, and showed that it has a universal amplitude predicted by the field theory. Our numerical simulations 
in dimension $d=1$ confirm this prediction. This may prove useful in analyzing finite-size corrections in experiments. 

We finally considered charge-density waves, which are related to loop-erased random walks and the $O(n)$-model at $n=-2$. We find that the amplitude of the critical force at depinning has a logarithmic dependence on the regularization scale, and that this can be understood in the framework of log-CFT. 

It would be interesting to also obtain  the
corrections to the dynamical exponent $z$, and we made some progress in this direction. 
The diagrams which   need to be evaluated    are much more involved, as sums of squared independent loop momenta appear in the denominator (see \Eq{second-line}), and the number of independent diagrams may well be a hundred. For this reason we decided to postpone their analysis to the future. 

With the 3-loop result at hand, another open question can   be tackled, namely the large-order behavior of  functional field theories, i.e.\ theories where the 
coupling constant is not a   number, but a function. We hope to report progress in this direction soon.

\acknowledgements
We thank Andrei Fedorenko, Jesper Jacobsen, Gauthier Mukherjee and Alberto Rosso for discussions.


\appendix

\section{Loop-Integrals}
\label{app:Integrals}
Here we give all loop integrals necessary for the main text. Some of them 
are calculated directly, while the remaining ones can be found in \cite{WieseHusemannLeDoussal2018}.

\subsection{The integral $I_{1}$}\label{app:I1}
The integral $I_{1}$ is defined as 
\begin{equation}\label{I1}
I_{1}= \Ione  := \int_{k} \frac{1}{(k^{2}+m^{2})^{2}}.
\end{equation}
It is calculated as follows:
\begin{eqnarray}\label{lf14}
I_{1}&=& \int_{k} \int_{0}^{\infty } \rmd \alpha \, \alpha \,\rme^{{-\alpha
(k^{2}+m^{2})} }\nn \\
 &=& \left(\int_{k}\rme^{-k^2} \right)  \int_{0}^{\infty
} \rmd \alpha\,  \alpha^{{1-\frac{d}{2}}} \, \rme^{-\alpha m^{2}}\nn \\
&=&  \left(\int_{k}\rme^{-k^2} \right) m^{-\epsilon } \Gamma
\left(\frac{\epsilon }{2} \right) .
\end{eqnarray}
This gives us the normalization-constant for higer-loop calculations
\begin{equation}
\left(\E I_1 \right) = m^{-\E}  \left(\int_{k}\rme^{-k^2} \right) \E
\Gamma \left(\frac{\epsilon }{2} \right) .
\end{equation}

\subsection{The tadpole diagram $I_{\rm tp}$}
\label{s:Itp}
Using that 
\be
-\frac{\partial}{\partial (m^2)}\, \ITP = \Ione
\ee
we get by integration that
\be
\label{diag:ITP}
\ITP = \frac{2
   m^{2}}{(d-4)
   (d-2)} (\epsilon I_1).
\ee

 \subsection{The integral $I_{A}$}\label{s:IA}
\begin{eqnarray} \label{IA}
\lefteqn{I_{A}= \ILH =}\nn \\
&=& 
\left[\frac{1}{2 \epsilon^2} + \frac{1}{4 \epsilon}+\frac{27{+}3 \psi' ({\textstyle 
\frac{5}{6}}){-}3\psi' ({\textstyle
\frac{1}{3}}) {-} 4 \pi ^{2}}{216} +\ca O (\epsilon )\right]\!
(\epsilon I_1)^2\nn \\
&=& \left[ \frac1{2\epsilon^2} + \frac1{4\epsilon} + \frac{1{-}4 C_3}{8}+\ca O(\epsilon)\right](\epsilon I_1)^2 
 \label{A.18}
\end{eqnarray} 
In the first line we gave the raw result obtained via computer algebra \cite{WieseHusemannLeDoussal2018}. 
The reflection
properties of the $\Gamma $-function combined with   the duplication and
triplication-formulas \cite{AbramowitzStegun}, give non-trivial relations, which following \cite{HusemannWiese2017,WieseHusemannLeDoussal2018}, are  used to combine all non-trivial terms into a single number, $C_3$:
\begin{eqnarray}
\textstyle 
\psi' (\frac{5}{6}) -\psi' (\frac{1}{3}) &=&\textstyle 4 \pi
^{2 }- 5 \psi' (\frac{1}{3})+ \psi' (\frac{2}{3}), \qquad
\label{psirel1} \quad 
\\\textstyle   
\psi' (\frac13) +\psi' (\frac{2}{3}) &=& \frac{4
\pi ^{2}}{3}, \label{psirel2}
\\\textstyle   
\psi' (\frac16) +\psi' (\frac56) &=& 4\pi ^{2} \label{psirel3}\\
C_3 =\frac{\psi'(\frac13)}6-\frac{\pi^2}9 &\approx& 0.585977.
\end{eqnarray}

\subsection{The sunset diagram}
\label{s:The sunset diagram}
\be
I_{\rm ss}= \Iss.
\ee
This diagram can be reduced to $I_A$ given in \Eq{A.18}. 
\be
-\frac{\partial}{\partial (m^2)}\, \Isunset = 3 \ILH.
\ee
Using \Eq{A.18} this implies
\be
I_{\rm ss} = m^2\left[-\frac{3}{2
   \epsilon ^2}-\frac{9}{4 \epsilon }-\frac{3(7-4 C_3)}{8}  + ... \right] (\epsilon I_1)^2. 
\ee

\subsection{The integral $I_{m}\equiv I_{o}$}\label{app:Im}\label{app:Io}
\bea
I_{m}&=&\IIm  
\equiv I_o = \Io  
 \nn \\
&=&  \bigg[ \frac{1}{3 \epsilon ^3}+\frac{1}{3 \epsilon ^2} +\frac{1-6 C_3}{6\epsilon}+\ca O (\epsilon  ) \bigg](\epsilon I_1)^3. \qquad 
\eea

\subsection{The star integral $I_{i}$ }\label{app:IM}
\begin{eqnarray}\label{Ii}
I_{i}= 
\Ii&=&\frac{\zeta(3)}{2\epsilon} \left ( \epsilon
I_{1}\right)^3 +\ca O (\epsilon^{0} ) \nn \\
&=&\frac{0.601028}{\epsilon } \left ( \epsilon
I_{1}\right)^3 +\ca O (\epsilon^{0} )\ . \qquad 
\end{eqnarray}

\subsection{The integral $I_{j}$}\label{app:Ij}

\begin{equation}
I_{j} = \Ij 
= \left[\frac{1}{3\epsilon
^{3}}+\frac{1}{6\epsilon ^{2}}+\frac{1}{12 \epsilon }+ \dots  \right]
(\epsilon I_{1})^{3} .
\end{equation}

\subsection{The integral $I_{l}$}\label{app:Il}
\bea
I_{l} &=& \Il  
\nn\\
&=& 
\left[ \frac{1}{6 \epsilon
   ^3}+\frac{1}{4 \epsilon
   ^2}+\frac{7-12 C_3}{24
   \epsilon }+\ca O (\epsilon) \right](\epsilon I_1)^3.\quad
\eea
 
\subsection{The integral $I_{r}$}
\label{a:Ir}
\be
I_r = \Ir.
\ee
This integral can be reduced to known integrals via a derivative w.r.t.\ $m^2$:
\bea
&&-\frac{\partial}{\partial (m^2)}~ \Ir = \Io  
+ 4\, \Il  
\nn\\
&& = \left[ \frac{1}{\epsilon ^3}+\frac{4}{3 \epsilon ^2} + \frac{\frac{4}{3}-3 C_3}{\epsilon}+\ca O(\epsilon^0) \right](\epsilon I_1)^3.
\eea
Integrating yields
\bea\label{Ir}
\!\!\! I_r&=& \Ir \nn\\
&=& m^{2} (\epsilon I_1)^3\left[-\frac{1}{\epsilon ^3}-\frac{17}{6 \epsilon
   ^2}+\frac{36 C_3-67}{12 \epsilon }+\ca O(\epsilon^0) \right] . \qquad 
\eea

\subsection{The integral $I_1$ for a finite system}
Define 
\bea
I_1^{\rm discrete}(m,L,d) &:=& \frac1{L^d} \sum_{\vec n} \frac1{\big[  (\frac {2\pi \vec n}{L}  ) ^2+m^2\big]^2}\nn\\
&=& \frac1{m^4 L^d} \sum_{\vec n} \frac1{\big[  (\frac {2\pi \vec n}{m L}  ) ^2+1\big]^2}. \qquad
\eea
If $mL\gg 1$, this can be approximated by an integral $\sum_{\vec n} \to \int \rmd^d \vec n$
\bea
I_1^{\rm discrete} (m,L,d)&=& \frac{(mL)^d}{m^4 L^d}\int \frac{\rmd^d \vec n}{(2\pi)^d}\frac1{( \vec n^2    +1)^2}\nn\\
&=& m^{-\epsilon} \int \frac{\rmd^d \vec n}{(2\pi)^d}\frac1{( \vec n^2    +1)^2}. 
\eea
This integral is given in \Eq{lf14}. 
On the other hand, 
\bea
&&\!\!\!I_1^{\rm discrete}(m,L,d)=  L^{4-d} \sum_{\vec n} \frac1{\big[  ( {2\pi \vec n}   ) ^2+(mL)^2\big]^2}\nn\\
&&=  L^{4-d} \int_{s>0} s \left[ \sum_{n=-\infty}^{\infty} \rme^{-s (2\pi n)^2 } \right] \rme^{-s (mL)^2}\nn\\
&&= L^{4-d} \int_{s>0}s\, \vartheta _3\Big(0,e^{-4 \pi ^2 s}\Big)^d \rme^{-s (mL)^2}
\eea
Therefore
\bea
\ca I_1(mL,d) &:=  & m^4 L^d I_1^{\rm discrete}(m,L,d) \nn\\
&&= (mL)^{4} \int_{s>0}s\, \vartheta _3\Big(0,e^{-4 \pi ^2 s}\Big)^d \rme^{-s (mL)^2}\nn\\
&&=   \int_{s>0}s\, \vartheta _3\Big(0,e^{-4 \pi ^2 s/(mL)^2}\Big)^d \rme^{-s}.
\eea
For small $mL$ one has
\be
\lim _{x\to 0} \ca I_1(x,d) = 1.
\ee
In the   limit of large $mL$ this gives
\bea
&& \ca I_1(mL,d) \simeq  \frac1{(4\pi)^{d/2}} \Gamma
    (  \textstyle \frac{4-d}{2} )  (mL)^d, \\
&&    \frac{\textstyle \Gamma
    (\frac{4-d}{2} ) }{(4\pi)^{d/2}} = \left\{ \begin{array}{ccc}
   \frac14 &\mbox{in} & d=1 \\
     \frac1{4\pi} &\mbox{in} & d=2 \\
       \frac1{8\pi} &\mbox{in} & d=3
   \end{array}\right. .
\eea

\section{Details for CDWs}

\subsection{$f_{\rm c}$ for CDWs in the limit of  $d\to 2$}
\label{CDW-d=2}
In $d=2$ we have (see e.g.~Eqs.~(100) and (101) of \cite{KompanietsWiese2019})
\be
\nu =\frac1{2-2 h_{1,3}} = 
\frac14\left(1+\frac \pi{\arccos(\frac n2)}\right) \ .
\ee
\bea
\eta &=& 4 h_{\frac 1 2,0}   
\nn\\
&=& \frac{5}{4} -\frac{3 \arccos \left(\frac{n}{2}\right)}{4 \pi
   }-\frac{\pi }{\arccos  \left(\frac{n}{2}\right)+\pi
   }. 
\eea
Sadly, 
\be
-\partial_n \left[ \frac1\nu +\eta\right]_{d=2} \sim \frac1{\sqrt{n+2}}
\ee
This may be related to the naturally appearing explicit factor of $1/(2-\epsilon)$.
The latter comes   when relating   diagrams for $f_{\rm c}$ to derivatives of known 
diagrams correcting the disorder. Undoing this integration then leads to a factor of 
$1/(2-\epsilon)=1/(d-2)$, see e.g.\ \Eq{diag:ITP}.

\subsection{The critical force as an observables inside the theory at $n=-2$}
\label{Observables inside the theory $n=-2$}

We find that \Eq{138} can   be calculated as follows in the theory at $n=-2$:
\be
-\frac{g}2 \left< \phi_2(x) \phi_1(x)^2  \rme^{-\frac g 8\int  (\vec\phi^2)^2}\right>_0\Big|_{n=-2} =  f_{\rm c}\, \phi_2 
\ee
In principle one should retain only 1PI diagrams, but this seems not to be necessary. The reason is probably that disconnected   and 1PR diagrams have additional factors of $(n+2)$.
The idea behind this contraction is to apply $\partial_n \phi^2$ to the interaction, which leads to something like $\phi_1^2 (\vec \phi^2)$. 
The ``additional component'' is represented by $\phi_1^2$; the problem is that it should not be equal to the other fields in the interaction. 
So the idea is to start constructing $\Gamma^{(2)}$ by selecting one external leg with component number 2, and then restricting the multiplying factor of $\vec \phi^2$ to a distinct component. The reason for   pulling out only one external (uncontracted) field is that otherwise we could either derive twice the same vertex or two vertices each once, which would complicate the writing.

An alternative formula is 
\bea
&&\!\!\!-\frac{g}8 \left< \Big[\phi_1(x) {-}\phi_2(x)\Big] \Big[ \phi_1(x)^2 {+}\phi_2(x)^2 \Big]\rme^{-\frac g 8\int  (\vec\phi^2)^2}\right>_0\Big|_{n=-2}
\nn\\
&& = f_{\rm c} (\phi_1-\phi_2)
\eea
I.e.\ we drop the space dependence as usual. 
Another alternative is
\be
-\frac g{6} \sum_i   \phi_i(x)^3\rme^{-{\ca S}} \to f_{\rm c}  \sum_i \phi_i 
\ee
Still another alternative is
\be
 {-\frac g{6}    \phi_1(x)^3 \rme^{-{\ca S}} \to f_{\rm c} \phi_1 }. 
\ee
This is given in the main text. 

\subsection{The critical force with complex fields}
This can also be done with $N$ complex fields in the limit of $N\to -1$. 
We use the action 
\be
\ca S = \int_x \nabla \vec \phi^*(x) \nabla \vec \phi(x) + m^2 \vec \phi^*(x)   \vec \phi(x)+\frac g 2 
\left[  \vec \phi^*(x)   \vec \phi(x\right]^2 . 
\ee
We find up to 4-loop order
\be
-\frac{g}2 \phi^*_1(x) \phi^*_2(x) \phi_2(x) \rme^{-\ca S}\Big|_{N=-1} \to f_{\rm c} \phi_1^* 
\ee
Another option is (gain checked up to 4-loop order)
\be
-\frac{g}4 \phi^*_1(x) \phi^*_1(x) \phi_1(x) \rme^{-\ca S} \Big|_{N=-1}= f_{\rm c} \phi_1^* .
\ee
The rational connecting these two observables is that
\bea
&-&\frac g2  \phi^*_1(x) \sum_i\phi^*_i(x) \phi_i(x) \rme^{-\ca S} \nn\\
&=&-\frac g2 \phi^*_1(x)  \phi^*_1(x) \phi_1(x) \rme^{-\ca S} \nn\\
&& -\frac g2 (N-1)  \phi^*_1(x)  \phi^*_2(x) \phi_2(x) ) \rme^{-\ca S}  \nn\\
&\to&  f_{\rm c} \phi_1^*  \left[ 2+   (N-1)\right] \to 0 \mbox{ at }N=-1.
\eea

\section{UV-cutoff dependent contributions to the critical force}
In the preceding sections, all diagrams were calculated within dimensional regularization, i.e.\ without an explicit UV cutoff. However, this is incorrect, as all diagrams have a strong UV-divergence. Here we wish to show that these additional UV-cutoff dependent terms are either independent of $m$, or at least  this dependence vanishes when we take $\Lambda$ large. 

There are two relatively simple ways to put an UV cutoff, 
\bea
I_{\rm tp}^{\text{hard}} &:=& \ITP = \int_k\frac1{k^2+m^2} \Theta(|k|\le \Lambda ),\\
I_{\rm tp}^{\text{soft}} &:=& \ITP = \int_k\frac1{k^2+m^2} \rme^{-a k^2 },\\
a &=& \frac 1 {\Lambda^2}, 
\eea
where $\Lambda$ is a large-momentum scale, of the same ingenering dimension as $m$. 
The soft cutoff gives 
\bea
\frac{I_{\rm tp}^{\text{soft}}}{\epsilon I_1|_{m=1}} &=& -\frac{\left(2-\frac{d}{2}\right)
   \Lambda^{d-2} e^{am^2}
   E_{\frac{d}{2}}( m^2/\Lambda^2)}{(d-4)
   \Gamma \left(3{-}\frac{d}{2}\right)}\nn\\
   &=& \frac{   \Lambda^{d-2}}{(d{-}2) \Gamma
    (3-\frac{d}{2} )}+\frac{2
   m^{{d}-2}}{(d{-}4)
   (d{-}2)} + \ca O(\Lambda^{-1}),\nn\\ 
\eea
where $E$ is the ``ExpIntegralE'' function. 

The hard cutoff gives
\bea\label{hardcutoff}
\frac{I_{\rm tp}^{\text{hard}}}{\epsilon I_1|_{m=1}} &=& -\frac{4 \Lambda ^d \sin (\frac{\pi 
   d}{2} ) \,
   _2F_1\Big(1,\frac{d}{2};\frac{d+2}{2}
   ;-\frac{\Lambda ^2}{m^2}\Big)}{\pi 
   (4-d) (d-2) d m^2} \nn\\
   &=& \frac{4 \Lambda ^{d-2} \sin
   \left(\frac{\pi  d}{2}\right)}{\pi 
   (d-4) (d-2)^2}+\frac{2 m^{d-2}}{(d-4)
   (d-2)}  + ... \qquad 
\eea
The strong UV divergence can   be extracted by applying a $\Lambda$ derivative,  
\bea
\Lambda\partial_\Lambda \frac{I_{\rm tp}^{\text{hard}}}{\epsilon I_1|_{m=1}} &=& \frac{4   \sin \left(\frac{\pi 
   d}{2}\right)}{\pi  (d-4) (d-2)}\times \frac{\Lambda^d}{m^2 +\Lambda^2}\nn\\
   &\sim& \Lambda^{d-2} + \ca O(m^2)  \Lambda^{d-4} .
\eea
The first term is $m$-independent, the second disappears in dimension $d<4$ for $\Lambda\to \infty$. 

Let us now apply this to the 2-loop sunset integral, 
\be
\Lambda\partial_\Lambda  I_{\rm ss} \sim \frac{3 \Lambda^d}{m^2 +\Lambda^2} \times \int_k \frac{{\Theta(|k|\le \Lambda)\Theta(|k+\Lambda|\le \Lambda)}}{(k^2+m^2) [(k+\Lambda)^2+m^2]}
\ee
The last factor has no IR singularity at $k\to 0$ or $k\to -\Lambda$. It can globally be bounded by $\Lambda^{d-4}$; for $k\to 0$ it goes as 
$m^{d-2}\Lambda^{-2}$. All these terms are IR finite in the limit of $\Lambda\to \infty$, $m\to 0$.

For the 3-loop integral, in a hard-cutoff scheme, 
\be
\Lambda \partial_\Lambda \Ir = \IrA + 4 \IrB, 
\ee
where the open circle indicates the momentum vector put to $\Lambda$. For the first, the momentum $\Lambda$ traverses both loops, s.t.
\be
\IrA = \mbox{IR-finite}.
\ee
Only the last diagram can give a contribution
\be
\IrB \simeq  {\Lambda^{-\epsilon}}  \times \frac{m^{-\epsilon}}{\epsilon} \times m^{d-2}.
\ee
We expect
the factor of $\frac{m^{-\epsilon}}{\epsilon}$ coming from the subdivergence in the lower loop to  be canceled by a counter term of the disorder.

\section{Critical force in $d=0$}
\label{Critical force in d=0}
In $d=0$, according to \cite{terBurgWiese2020}
\bea
f_{\rm c} &\simeq& \sqrt{2 \ln (m^{-2})} +\frac{\gamma_{\rm E}}{\sqrt{2 \ln (m^{-2})}}+...
\qquad \mbox{(70) of \cite{terBurgWiese2020}}  \nn \\
\rho_m &\simeq& \frac1{m^2 \sqrt{2 \ln (m^{-2})}} \quad \mbox{(58) of \cite{terBurgWiese2020}}\qquad\\
\Delta (w) &=& m^4 \rho_m^2 \tilde \Delta(w/\rho_m) \quad \mbox{(60) of \cite{terBurgWiese2020}}\qquad \nn
\eea
This gives in $d=0$ the two combinations  of the main text, 
\bea
\tilde {\ca A} m^2 \rho_m &=&  \frac{\tilde {\ca A}}{\sqrt{ 2 \ln (m^{-2})}} \\
{|f_{\rm c}|} &=&  \sqrt{ 2 \ln (m^{-2})} + \frac{\gamma_{\rm E}}{\sqrt{ 2 \ln (m^{-2})}}  + ...\qquad 
\eea
To our disappointment, the singularities of these two terms are different, so that we cannot obtain the amplitude $\tilde {\ca A}$ in $d=0$.

\enlargethispage{1cm}

\section{A worked-out  example: logarithmic operators for self-avoiding polymers}
\label{s:Logarithmic operators for self-avoiding polymers: Considerations from Field Theory} 
Following Cardy  \cite{Cardy1999}, (see \cite{Cardy2013} for an extended review), we consider the logarithms appearing for self-avoiding polymers.
To this aim, 
introduce the polymer density in the $\phi^4$ field theory for polymers \cite{DeGennes1972}, which transforms as a singlet under  ${O}(n)$\footnote{Contrary to the conventions  Cardy uses in  \cite{Cardy1999} we divided $\ca E$ by $n$ to simplify notations. These  are the   conventions he  later  uses in \cite{Cardy2013}.}, 
\be
{\cal E}_{i} :=\phi_{i}^{2}\ , \qquad 
{\cal E}:= {  \frac 1 n}\sum_{i=1}^{n} \phi_{i}^{2}  \ .
\ee
Next consider the traceless vector 
\be
\tilde{\mathcal E}_{i} := \phi_{i}^{2}-\frac1n \sum_{j=1}^{n} \phi_{j}^{2} \equiv {\cal E}_{i} -{\cal E}\ .
\ee
Alternatively one can use the traceless  tensor operator, which sits   in the same multiplet
\be
\tilde {\cal E} _{{ij}}:= \phi_{i}\phi_{j} - \frac1n \delta_{ij} \sum_{k=1}^{n}\phi_{k}^{2}\ .
\ee
In these notations, 
\bea
x_{\cal E}(n)=\mbox{dim}_{\mu}({\cal E}) &=& d-y_{\cal E}= d- 2 - \gamma_{{\phi^{2}}}+ \eta, \qquad \\
x_{\tilde {\cal E}}(n)=\mbox{dim}_{\mu}(\tilde {\cal E}) &=& d-y_{\tilde {\cal E}}=d- 2 -\gamma_{{\phi\phi}}+ \eta. 
\eea
Then 
\bea
\left< {\cal E} (r) {\cal E}(0) \right> &=& {  \frac 1 n} \Big[ \left< {\cal E}_{1} (r) {\cal E}_{1}(0) \right> +(n-1) \left< {\cal E}_{1} (r) {\cal E}_{2}(0) \right>\Big] \nn\\
&\simeq& { \frac {A(n)} {  n}}   r^{{-2 x_{\cal E} (n)}},
\eea
\bea
\left< \tilde {\cal E}_{i} (r) \tilde{\cal E}_{i}(0) \right> &=& \frac{n-1}n \Big[ \left< {\cal E}_{1} (r) {\cal E}_{1}(0) \right> - \left< {\cal E}_{1} (r) {\cal E}_{2}(0) \right>\Big]\nn\\
& \simeq &\frac{n-1}n   \tilde A(n) r^{{-2  x_{\tilde{\cal E}} (n)}}. 
\eea
Since the expressions in the square brackets become identical in the limit of $n\to 0$, 
\be
A(0) = \tilde A(0)\ , \qquad x_{\cal E}(0) = x_{\tilde {\cal E} }(0)\ .
\ee
Consider
\bea
&&  \left< {\cal E} (r) {\cal E}(0) \right> +  \left< \tilde {\cal E}_{i} (r) \tilde{\cal E}_{i}(0) \right> =   \left< {\cal E}_{1} (r) {\cal E}_{1}(0) \right> , \\
&& \left< {\cal E} (r) {\cal E}(0) \right> - \frac{1}{n-1} \left< \tilde {\cal E}_{i} (r) \tilde{\cal E}_{i}(0) \right> =   \left< {\cal E}_{1} (r) {\cal E}_{2}(0) \right>. \qquad 
\eea
This implies that 
\begin{widetext}
\bea
\left< {\cal E}_{1} (r) {\cal E}_{2}(0) \right> &=& \frac1n \Big[ A(n) r^{{-2 x_{\cal E} (n)}}-  \tilde A(n) r^{{-2  x_{\tilde{\cal E}} (n)}} \Big] \nn\\
&=&  A(n) r^{{-2 x_{\cal E} (n)}} \frac1n \Big[ 1-  \frac{\tilde A(n)}{A(n) } r^{{2[x_{\cal E} (n)- \tilde x_{\cal E} (n)]}} \Big]\nn\\
&=& A(0) r^{{-2 x_{\cal E} (n)}}\Big[ \frac{A'(0)-\tilde A'(0)}{A(0)}+2\ln(r)\Big( x_{\tilde {\cal E}}'(0)-x_{{\cal E}}'(0)\Big) \Big] + {\cal O}(n), 
\eea
\bea
\left< {\cal E}_{1} (r) {\cal E}_{1}(0) \right> &-& \left< {\cal E}_{1} (r) {\cal E}_{2}(0) \right> = \frac{n}{n-1}  \left< \tilde {\cal E}_{i} (r) \tilde{\cal E}_{i}(0) \right> = r^{{-2 x_{\tilde {\cal E}}(n)}} \tilde A(n), \\
\label{133}
 \left< {\cal E}_{1} (r) {\cal E}_{1}(0) \right> 
 &=&A(0) r^{{-2 x_{\cal E} (n)}}\Big[1+ \frac{A'(0)-\tilde A'(0)}{A(0)}+2\ln(r)\Big( x_{\tilde {\cal E}}'(0)-x_{{\cal E}}'(0)\Big) \Big] + {\cal O}(n).
\eea
\end{widetext}
As a consequence,  the ratio reads
\bea
&&\frac{\left< {\cal E}_{1} (r) {\cal E}_{2}(0) \right>}{ \left< {\cal E}_{1} (r) {\cal E}_{1}(0) \right> - {\left< {\cal E}_{1} (r) {\cal E}_{2}(0) \right>}} \nn\\
&&~ = \frac{A'(0){-}\tilde A'(0)}{A(0)}+2\ln(r)\Big( x_{\tilde {\cal E}}'(0){-}x_{{\cal E}}'(0)\Big)+ \ca O(n).\qquad ~~
\eea
Denoting by a colored circle a self-avoiding polymer, the l.h.s.\ can be written as
\be
\frac{\left< {\cal E}_{1} (x) {\cal E}_{2}(y) \right>}{ \left< {\cal E}_{1} (x) {\cal E}_{1}(y) \right> - {\left< {\cal E}_{1} (x) {\cal E}_{2}(y) \right>}}
=\frac12~
\frac{\parbox{2.5cm}{{\begin{tikzpicture}
\node (x) at  (0.5,0.5)    {$\parbox{0mm}{$\raisebox{-1mm}[0mm][0mm]{\!\!\!\!$\scriptstyle x$}$}$};
\node (y) at  (2.75,0.5)    {$\parbox{0mm}{$\raisebox{-1mm}[0mm][0mm]{$\,\,\scriptstyle y$}$}$};
\fill (x) circle (1.5pt);
\fill (y) circle (1.5pt);
\draw [blue,thick](0.5,0.5) arc (-180:180:0.5);
\draw [red,thick](1.75,0.5) arc (-180:180:0.5);
\end{tikzpicture}}}}
{\parbox{2.5cm}{{\begin{tikzpicture}
\node (x) at  (0.5,0.5)    {$\parbox{0mm}{$\raisebox{-1mm}[0mm][0mm]{\!\!\!\!$\scriptstyle x$}$}$};
\node (y) at  (2.75,0.5)    {$\parbox{0mm}{$\raisebox{-1mm}[0mm][0mm]{$\,\,\scriptstyle y$}$}$};
\fill (x) circle (1.5pt);
\fill (y) circle (1.5pt);
\draw [blue,thick](0.5,0.5) arc (-120:-60:2.25);
\draw [red,thick](0.5,0.5) arc (120:60:2.25);
\end{tikzpicture}}}} \ .
\ee
The numerator is the probability that two ring-polymers attached at $x$ and $y$ do not intersect. The denominator is the probability that the ends of two  polymers attached  at $x$ and $y$ are at a distance $x-y$. 
According to \Eq{133} this ratio contains a logarithmic contribution, with a universal amplitude given by the derivatives of the critical exponents. 
Explicit numerical values are given in Ref.~\cite{KompanietsWiese2019}. 

Let us finally introduce the logarithmic pair. 
Following   Cardy \cite{Cardy2013},   define  
in the limit of $n\to 0$, 
\bea
{\cal C} &:=&\lim_{{n\to 0}} [x_{\ca E}(n)-x_{\tilde{\ca E}}(n)]\ca E  \nn\\
&\equiv& \lim_{{n\to 0}} [x_{\ca E}(n)-x_{\tilde{\ca E}}(n)]\tilde {\ca E} \qquad  \\
{\cal D} &:=& \lim_{{n\to 0}} \ca E - \tilde {\ca E}
\eea
This implies 
\bea
\!\!\!\!\left< \ca D(0) \ca D(r) \right> &=&\lim_{{n\to 0}} \frac1n \Big[ A(n) r^{{-2 x_{\cal E} (n)}}-  \tilde A(n) r^{{-2  x_{\tilde{\cal E}} (n)}} \Big] \nn\\
& =& -\frac{-2 \alpha \ln (r) + \mbox{const}}{r^{{2 x(0)}}} , \\
\!\!\!\!\left< \ca C(0) \ca D(r) \right> &=& \lim_{{n\to 0}} [x_{\ca E}(n)-x_{\tilde{\ca E}}(n)]\left< \ca E (0) [\ca E(r) -\tilde {\ca E}(r)]\right> \nn\\
&   =& \frac{ \alpha}{r^{{2 x(0)}}}  , \\
\left< \ca C(0) \ca C(r) \right> &=& \lim_{{n\to 0}} [x_{\ca E}(n){-}x_{\tilde{\ca E}}(n)]^{2}\left< \ca E (0)  \ca E(r)  \right>  = 0\ . \\
\alpha &=&  A(0)     \Big(x_{{\cal E}}'(0)- x_{\tilde {\cal E}}'(0)\Big)\quad \nn\\ 
&\equiv&  \tilde A(0)     \Big(x_{{\cal E}}'(0)- x_{\tilde {\cal E}}'(0)\Big)\ .
\eea
$(\ca C, \ca D)$ forms a logarithmic pair. Denoting the dilation operator by $\mathbb D$,  away from the point of degeneracy $n=0$, 
\bea
\mathbb D \circ \ca E &=& x_{\ca E}(n)\, \ca E , \\
\mathbb D \circ \tilde {\ca E} &=& x_{\tilde {\ca E}}(n)\, \tilde {\ca E}. 
\eea
This implies with $x:= x_{\ca E}(0) \equiv x_{\tilde {\ca E}}(0)$ 
\bea
\mathbb D \circ \ca C &=& x \, \ca C, \\
\nn
\mathbb D  \circ   {\ca D} &=& \lim_{n\to 0}  x_{\ca E}(n)  \ca E - x_{\tilde {\ca E}} (n)\tilde {\ca E} \\
&=& \lim_{n\to 0}  [x_{\ca E}(n) - x_{\tilde {\ca E}}(n) ]  \ca E + x_{\tilde {\ca E}} (n)[   {\ca E} -\tilde {\ca E} ] \nn \\
&=& \ca C + x \ca D.
\eea
Written in matrix form, the dilatation operator has a (non-diagonalizable) block-Jordan form, 
\be
{\mathbb D}\circ \bigg( { \ca C \atop \ca D}\bigg)  = {\textstyle \bigg( \textstyle
\begin{array}{cc}  
\textstyle
x & \textstyle 0 \\
 \textstyle 1 &\textstyle x
\end{array}\bigg) } \bigg( { \ca C \atop \ca D} \bigg) . 
\ee
\begin{figure}[t]    
\includegraphics[width=\columnwidth]{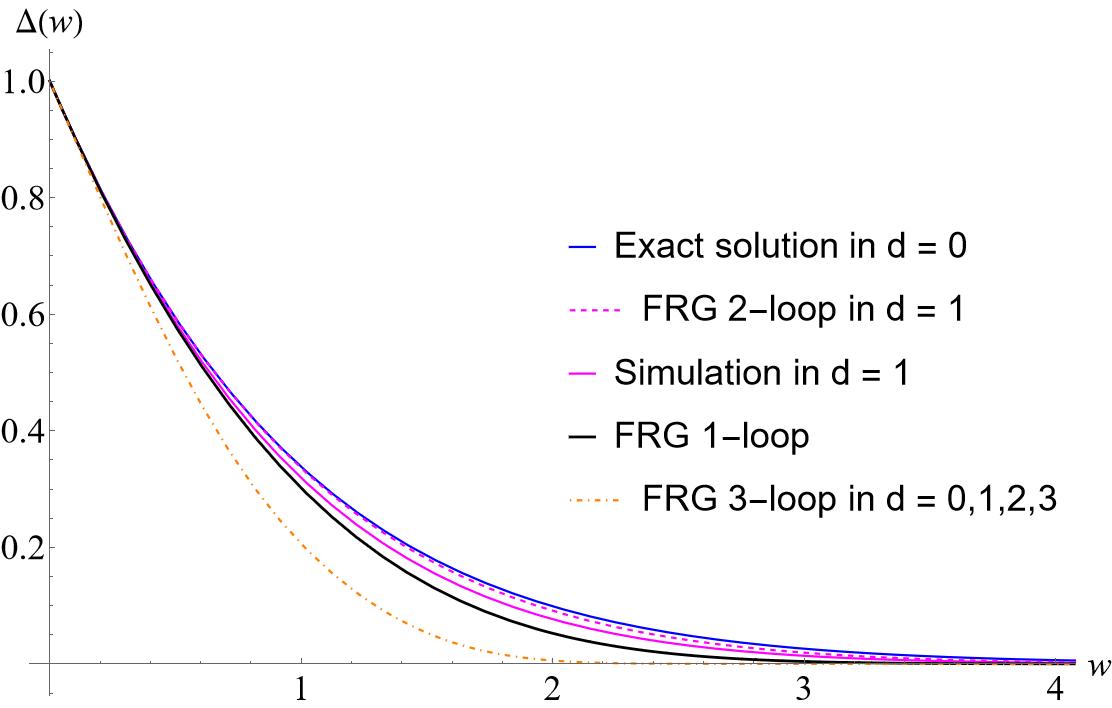}
\caption{Shape comparison of 1-loop FRG result (black, thick, solid), improved 2-loop FRG result as given by Eq.~\ref{Pade-Delta} (magenta dashed), simulation in    dimension $d=1$ (magenta), exact solution in dimension $d=0$ (blue), and 3-loop improved FRG results as given by Eq.~\ref{rescaled-Delta-3} for   values of dimension $d=0, 1, ..., 3$. The last four curves (orange, dot-dashed) are indistinguishable.}
\label{fig:f1}
\end{figure}
\section{Improvment of 3-loop result}
\label{f:3-loop resumation}

There are two improvements we tried in our comparison between theory and simulations: The first is a Pad\'e-resummation, as in \Eq{Pade-Delta}, continued to 3-loop order. This strategy failed. 

Our second attempt at improvement consisted in replacing 
\be\label{rescaled-Delta-3}
\Delta(w) \to \lambda^{-2} \Delta(w \lambda), \quad \lambda = 1 + \alpha \epsilon + \beta \epsilon^2+ \ca O(\epsilon^3).
\ee 
This transformation is an exact property of the RG equation. 
 We then  Taylor-expand \Eq{rescaled-Delta-3} to order $\epsilon^3$, and drop the higher-order terms. 
Let us stress that there is no natural choice for $\lambda$:   or choice of setting $\Delta(0)\stackrel!=\epsilon/3$, forces higher-order corrections to vanish at $w=0$. It is   one particular choice,   maybe not the best. 
 This procedure helps us enforce some physical properties of $\Delta(w)$, the most important one being that it has its maximum at  $w = 0$, and then decays linearly for small $w$. We succeeded to achieve this,  but we were unable to tune the   $\alpha$ and $\beta$ in order to get close to the analytical solution of \cite{LeDoussalWiese2008a} in dimension $d=0$,   or our simulation results in dimension $d=1$. Moreover, whenever we achieved a monotonic decay around $w=0$, the result for $\Delta(w)$ achieved by this transformation does not seem to depend on the dimension $d$, and the resulting curves lie way beyond the 1-loop curve as can be seen on Fig.~\ref{fig:f1}.

\section{Independence of $\mathcal{B}$ on the elastic coefficient $c$}
\label{g:B and c}

\begin{table}[t]
	\begin{tabular}{ |c| c| c|c|c| }
		\hline
		$c$ & 0.5 & 1 & 2 & 4\\ 
		\hline
		$f_c$ & $1.14 \pm 0.05$ & $0.97 \pm 0.05$ & $0.87 \pm 0.05$ & $0.78 \pm 0.05$\\  
		\hline
		$\hat{\rho}$ & $0.57 \pm 0.05$ & $0.54 \pm 0.05$ & $0.44 \pm 0.05$ & $0.42 \pm 0.05$ \\  
		\hline
		$\mathcal{B}$ & $2 \pm 0.2$ & $1.8 \pm 0.2$ & $1.98 \pm 0.2$ & $1.86 \pm 0.2$\\
		\hline
	\end{tabular}
\caption{Simulations results    for $\hat \rho := \rho_m m^{\zeta}$ (section \ref{s:Critical force}) and $f_c$ for different values of $c$, $L=128$, $mL/\sqrt{c}=4$.}
\label{tab:g1} 
\end{table}

\begin{figure}[t]
\includegraphics[width=\columnwidth]{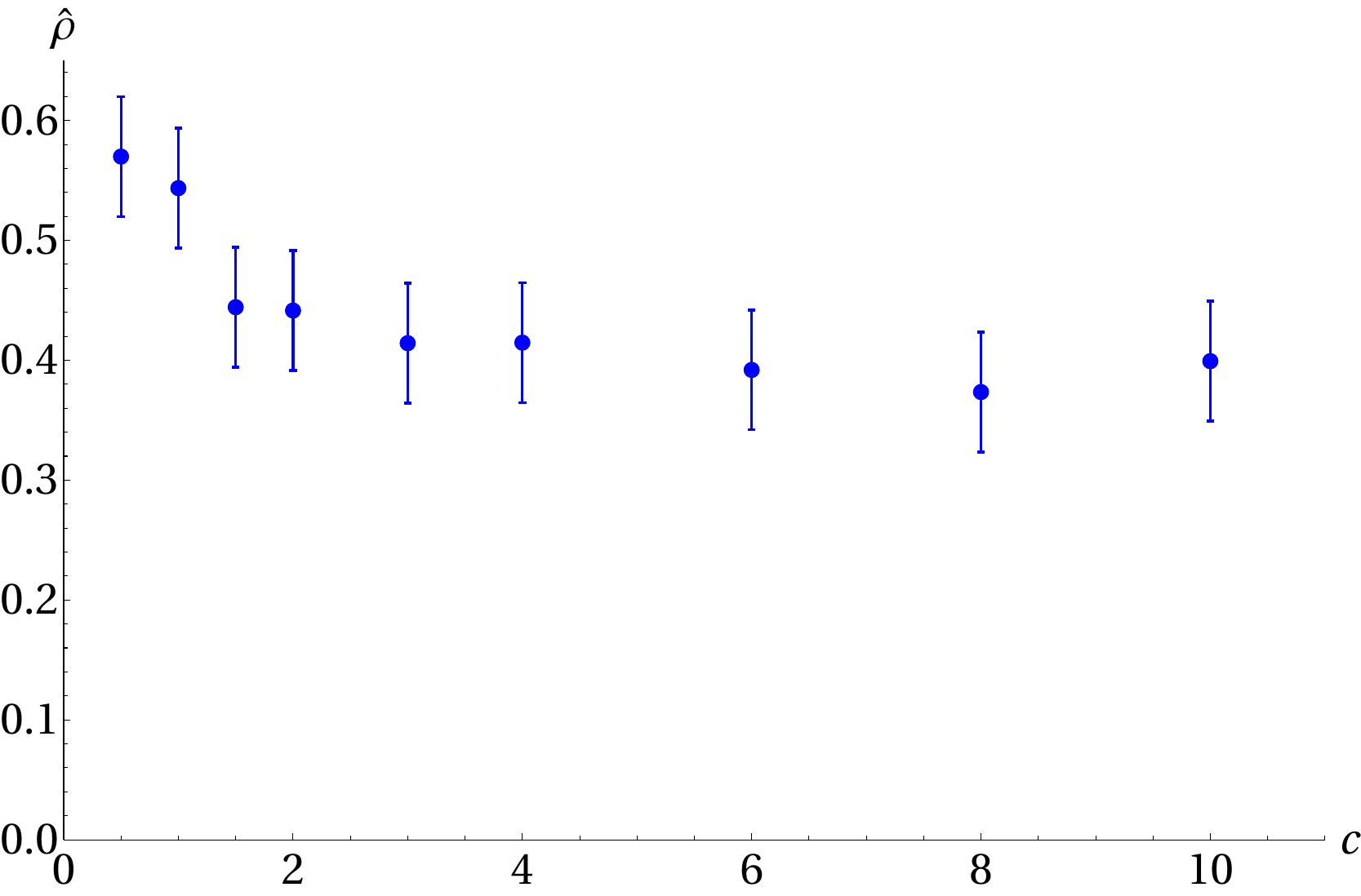}
\caption{Dependence of $\hat{\rho} $ on $c$.}
\label{fig:g1}
\end{figure}
To demonstrate that  the universal amplitude $\mathcal{B}$  defined in \Eq{amplitudeB} is independent of the elastic coefficient $c$, we perform simulations for different values of $c$. The results are presented in   table~\ref{tab:g1} and Fig.~\ref{fig:g1}.
Within error bars,  $\mathcal{B}$ does not change with   $c$.  Fig.~\ref{fig:g1} shows the   dependence of  the rescaled correlation length  $\hat{\rho}:= \rho_m m^{\zeta}$
on $c$. We observe an increase of $\hat \rho$ for $c<1$, which can be explained as follows. Making $c$ and $m$ smaller (we fix $mL = 4 \sqrt c$) renders the interface fluctuations larger, allowing it to explore more disorder configurations. As a consequence, $\hat \rho$ slightly increases. This shows numerically that while $\cal B$ is universal, $\hat \rho$ is not.

\setcounter{section}{17} 


\ifx\doi\undefined
\providecommand{\doi}[2]{\href{http://dx.doi.org/#1}{#2}}
\else
\renewcommand{\doi}[2]{\href{http://dx.doi.org/#1}{#2}}
\fi
\providecommand{\link}[2]{\href{#1}{#2}}
\providecommand{\arxiv}[1]{\href{http://arxiv.org/abs/#1}{#1}}
\providecommand{\hal}[1]{\href{https://hal.archives-ouvertes.fr/hal-#1}{hal-#1}}
\providecommand{\mrnumber}[1]{\href{https://mathscinet.ams.org/mathscinet/search/publdoc.html?pg1=MR&s1=#1&loc=fromreflist}{MR#1}}

\tableofcontents
\label{tableofcontents}

\end{document}